\documentclass{aa}  

\usepackage{graphicx}
\usepackage{xcolor}
\usepackage{mhchem}
\usepackage{txfonts}
\providecommand{\e}[1]{\ensuremath{\times 10^{#1}}}
\usepackage{hyperref}
\begin{document}

   \title{Radical Addition and \ce{H} Abstraction Reactions in \ce{C2H2}, \ce{C2H4} and \ce{C2H6}: A Gateway for Ethyl and Vinyl Bearing Molecules in the Interstellar Medium} 
   \titlerunning{Radical addition and \ce{H} abstraction reactions in \ce{C2H2}, \ce{C2H4} and \ce{C2H6}}

   \author{G. Molpeceres
          \inst{1}
          \and
          V.M. Rivilla
          \inst{2}
          }
          
   \institute{Institute for Theoretical Chemistry, University of Stuttgart, Stuttgart, Germany\\
              \email{molpeceres@theochem.uni-stuttgart.de}
         \and
            Centro de Astrobiología (CSIC-INTA), Ctra. de Ajalvir Km. 4, Torrejón de Ardoz, 28850 Madrid, Spain
             }

   \date{Received \today, accepted \today}

 
  \abstract
   { Recent interstellar detections include a significant number of molecules containing vinyl (\ce{C2H3}) and ethyl (\ce{C2H5}) groups in their structure. For several of these molecules, there is not a clear experimental or theoretical evidence that support their formation from simpler precursors}
   {We carried out a systematic search of viable reactions starting from closed shell hydrocarbons containing two carbon atoms (ethane, \ce{C2H6}; ethylene, \ce{C2H4}; and acetylene, \ce{C2H2}) with the goal of determining viable chemical routes for the formation of vinyl and ethyl molecules on top of interstellar dust grains.}
   {We used density functional theory calculations in combination with semiclassical instanton theory to derive the rate coefficients for the radical-neutral surface reactions. The effect of a surface was modelled through an implicit surface approach, profiting from the weak interaction between the considered hydrocarbons and the dust surfaces.}
   {Our results show that both H and OH radicals are key in converting acetylene and ethylene into more complex radicals that are susceptible to continue reacting and forming interstellar complex organic molecules. The relevant reactions, for example OH additions, present rate constants above 10$^{1}$ s$^{-1}$ that are likely competitive with OH diffusion on grains. Similarly, H atom addition to acetylene and ethylene is a very fast process with rate constants above 10$^{4}$ s$^{-1}$ in all cases, and greatly enhanced by quantum tunneling. Hydrogen abstraction reactions are less relevant, but may play a role in specific cases involving the OH radical. Reactions with other radicals \ce{NH2, CH3} are likely to have a much lesser impact in the chemistry of ethyl and vinyl bearing molecules. }
   {The effective formation at low temperatures of four radicals (\ce{C2H3}, \ce{C2H5}, \ce{C2H2OH}, and \ce{C2H4OH}) through our proposed mechanism opens the gate for the formation of complex organic molecules and indicates a potential prevalence of OH bearing molecules on the grain. Following our suggested reaction pathway, we explain the formation of many of the newly detected molecules, and propose new molecules for detection. Our results reinforce the recent view on the importance of the OH radical in the interstellar surface chemistry.  }


   \maketitle
%

\section{Introduction} \label{sec:introduction}

The definition of interstellar complex organic molecules (COMs) encompasses to all carbon-bearing molecules containing six atoms or more in their molecular backbone \citep{Herbst2009}. Under this broad definition fall many of the recently detected molecules in the interstellar medium (ISM). In the last years, the rate of detections has greatly increased \citep{Mcguire2022}. Under the cold conditions of molecular clouds and prestellar cores, where the recent surge of molecular detections has taken place, chemistry is significantly hindered. For example, all reactions that proceed through endothermic pathways or present prohibitively large activation barriers are precluded, conditioning the chemistry both in the gas phase (see \citet{Puzzarini2022} for a recent review) and on top of ice coated dust grains (\citet{Hama2013, Cuppen2017a}). Yet, with these restraints, evermore COMs are known to be present in the ISM. Ethyl and vinyl bearing molecules, e.g. molecules whose carbon skeleton backbone can be identified to proceed from either the vinyl (\ce{C2H3}) or ethyl (\ce{C2H5}) radicals have been particularly relevant, due to their prevalence, \ce{C2H5OH} (\citet{Zuckerman1975}), \ce{C2H3OH} (\citet{Agundez2021}),
\ce{C2H5CN} (\citet{Johnson1977}),
\ce{C2H3CN}  (\citet{Gardner1975}), 
\ce{C2H5OCHO} (\citet{Belloche2009}), \ce{C2H5OCH3} (\citet{Tercero2018}), \ce{C2H5CHO} and \ce{C2H3CHO} (\citet{Hollis2004}), \ce{C2H5SH} \citep{RodriguezAlmeida2021}, \ce{C2H5NCO} \citep{Rodriguez-Almeida2021b},
\ce{C2H3NH2} \citep{Zeng2021}, 
\ce{C2H3CCCH} \citep{Cernicharo2021c}, 
\ce{C2H3C3N} \citep{KelvinLee2021},
\ce{NH2C2H4OH} \citep{Rivilla2021}, or \ce{HOC2H2OH} \citep{Rivilla2022}. We emphasize from this list of molecules, that many of them have been detected in the last four years. 

The chemistry occurring on top of icy dust grains possess its particular idiosyncrasy, with weakly bound adsorbates landing on the grain, diffusing, reacting and eventually returning to the gas phase, where they are detected by radio telescopes. All these motions favor hydrogen atoms as main chemistry initiators on grains, owing to a fast diffusion \citep{Hama2012b, Senevirathne2017, asg17, Nyman2021} and the possibility of tunnel through potential energy barriers \citep{Hama2013}. However, and thanks to \textit{non-diffusive} mechanisms, appearing as a consequence of an excess of energy inoculated into a particle (i.e., after photodissociation, chemical reaction or cosmic-ray interaction \citet{Jin2020, Garrod2022}), other radicals (OH, \ce{NH2}, \ce{CH3}) may play an important role in the chemical evolution of the ISM \citep{Fedoseev2017,Chuang2020, Ioppolo2020, Tsuge2021, Ishibashi2021}.

In this work, we merge the knowledge attained by observations and experiments seeking for a mechanism able to explain the prevalence of ethyl and vinyl bearing molecules in cold environments. The mechanism that we propose, that does not exclude other possible gas-phase and surface scenarios, relies in the processing of the closed shell hydrocarbons acetylene (\ce{C2H2}), ethylene (\ce{C2H4}), and ethane (\ce{C2H6}) by reactive radicals through addition and H abstraction reactions. Either of these reactions forms a reactive radical that is prone to continue reacting with other incoming particles to form stable COMs. In particular, in this work we have considered addition and abstraction reactions of the above mentioned molecules with the following radicals: \ce{H}, \ce{OH}, \ce{NH2}, and \ce{CH3}. We employed a theoretical approach in our work, and as we emphasize throughout the paper, our predictions will significantly benefit from experimental validation.

The paper is divided as follows: in section \ref{sec:methods} we discuss the particularities of the computational framework to study the wide range of reactions under consideration, in section \ref{sec:results} we present our results, divided by parent closed-shell molecule. Finally in section \ref{sec:discussion} we gather all the trends obtained from the reactions, discuss the implications of out findings and postulate new chemical species susceptible of detection following the mechanism that we propose.

%

\section{Methodology} \label{sec:methods}

The computational setting for this work follows similar protocols that some other works by us \citep{Miksch2021}. Briefly, we have determined the activation energies ($\Delta U_{\text{a}}$, including zero-point energy contributions) and reaction energies ($\Delta U_{\text{e}}$) for a set of reactions of significance for the formation of ethyl and vinyl bearing radicals and starting from (\textbf{a}) acetylene (\ce{C2H2}), (\textbf{b}) ethylene (\ce{C2H4}), and (\textbf{c}) ethane (\ce{C2H6}). The reactions that we have considered for (\textbf{a}) are:

\begin{align}
    \ce{C2H2 / C2H4 + X &-> XC2H2 / XC2H4}, \label{eq:1}\\
    \ce{C2H2 / C2H4 + X &-> C2H / C2H3 + HX}, \label{eq:2}
\end{align}

\noindent with X= \ce{H}, \ce{OH}, \ce{CH3} and \ce{NH2}. Therefore, for \textbf{a,b} we have studied both addition reactions and abstraction ones. For (\textbf{c}), on the contrary, we have studied only \ce{H} abstraction reactions:

\begin{equation}
    \ce{C2H6 + X -> C2H5 + HX}, \label{eq:3}\\
\end{equation}

\noindent because additions to \textbf{b} would incur on a very high barrier as a consequence of ethane being completely hydrogen saturated. All the reactions were studied in the context of surface science aiming for an implementation of our values into astrochemical gas-grain surface models. Because of the weak binding of \textbf{a} and \textbf{b} with amorphous solid water (ASW), and presumably with other interstellar relevant surfaces \citep{Wakelam2017b}, all the reactions studied here account for the effect of a surface \emph{via} the implicit surface approach \citep{Meisner2017}, \emph{e.g} the rotational partition function of the transition state and the reactant is kept fixed in subsequent rate constant calculations (see below). This approach effectively considers that the reactions are hardly affected by the interaction with the surface and that the surface plays a role in increasing the concentration of reactants, as well as in dissipating the reaction energy for very exothermic reactions. Under this approach, the surface composition does not play any role, and therefore it should be assumed with care for other surfaces where the adsorbate-surface interaction is stronger. We tested the validity of the implicit surface approach for three reactions of our whole set of reactions in Appendix \ref{sec:appendix}. We consider unimolecular reactions in the context of a Langmuir-Hinshelwood mechanism, characteristic of surface reactions with a barrier. 

We studied the reactions with density functional theory (DFT) calculations. In this work, we use the MN15-D3BJ/def2-TZVP \citep{Weigend2005, Grimme2011,Yu2016} exchange and correlation functional.\footnote{The parameters for the dispersion correction term are the recommended ones extracted from \cite{Goerigk2017}, Supplementary Information, Table S.3. s$_{6}$=1.0, a$_{1}$=2.0971, s$_{8}$=0.7862 and a$_{2}$=7.5923} We chose this level of theory according to its performance in generally predicting activation barriers \citep{Yu2016} when comparing with refined energies at the coupled cluster level (UCCSD(T)-F12/cc-pVTZ-F12//MN15-D3BJ/def2-TZVP) values, considered as reference. All the energetic quantities presented in this work are given at the pure DFT level and with energies corrected using explicitly correlated coupled cluster theory mentioned above \citep{Knizia2009}, denoted in square brackets throughout the paper. For specific reactions of our study, coupled cluster theories are known to underperform against the values in the literature \citep{Senosiain2005, Senosiain2006}. These cases are explicitly indicated when relevant.  All the calculations employ Gaussian16 \citep{g16} interfaced to DL-Find/Chemshell \citep{Chemshell, kae09a,Quan2016}.

Once the stationary points in the potential energy surface (PES) are determined (reactant, transition states and products) we calculate the corresponding rate constants of reaction attending in their classical transition state formulation and incorporating quantum effects by means of semiclassical instanton theory \citep{lan67,mil75,col77,Rommel2011-2, Rommel2011}. The latter are determined below the crossover temperature using conventional instanton theory and reduced instanton theory \citep{McConnell2017} above it. Crossover temperatures ($T_{c}$) are defined as the temperatures below which quantum effects dominates over thermal effects and they are defined as in our previous works \citep{Miksch2021, Molpeceres2021, molpeceres2021diastereoselective}, following the formulation of \cite{Gillan1987}:

\begin{equation}
    T_\text{c} = \frac{\hbar \omega_{i}  }{2\pi k_\text{B}} ,
\end{equation}

\noindent where $\omega_{i}$ is the absolute value of the imaginary frequency for the activated complex and k$_\text{B}$ is the Boltzmann constant. Symmetry numbers ($\sigma$) are not included as pre-factors in our calculated rate constants because there is not a univocal symmetry number for a reaction on a surface due to the total or partial break of symmetry in such reactions. 

The anharmonicity of the instanton structures in \ce{C2H6-CH3} and \ce{C2H6-NH2} activated complexes precludes the calculation of reliable instanton rate constants at this stage of research, and this will be discussed later in the paper. Similarly, in \ce{C2H4 + OH -> C2H4OH} we could not converge instanton paths below 50~K. For all these exceptions, we calculated tunneling contributions stemming from a symmetric Eckart barrier to the rate constants. The latter are less exact than instanton calculations and should serve as an approximation to the real values. 

Reaction energies are provided, with reactant energies coming from the pre-reactive complex identified in intrinsic reaction coordinate calculations (IRC). The energies of the product are always computed for the separated products. We have created a \texttt{Zenodo} repository to store all the structures gathered in this article \citep{molpeceres_2022_6581077}. \footnote{https://zenodo.org/record/6581077\#.Yo96uRNBxhE}


\section{Results} \label{sec:results}

Throughout this section and section \ref{sec:discussion} the naming convention for the reactions will be AD/AB X.Y with AD=addition, AB=abstraction; X=1 (\ce{C2H2}), 2 (\ce{C2H4}) and 3 (\ce{C2H6}); Y=1 (H), 2 (OH), 3 (\ce{NH2}) and 4 (\ce{CH3}). 

\subsection{Radicals reacting with acetylene}

\begin{table}
\caption{Reaction energies ($\Delta U_{r}$ in kJ mol$^{-1}$)  and activation energies ($\Delta U_{a}$ in kJ mol$^{-1}$) for the addition and abstraction reactions starting from \ce{C2H2} at the MN15-D3BJ/def2-TZVP level. In square-brackets, values at the UCCSD(T)-F12/cc-pVTZ-F12//MN15-D3BJ/def2-TZVP level of theory.}             
\label{tab:c2h2a} 
\centering                          
\resizebox{\linewidth}{!}{
\begin{tabular}{c c c c }        
\hline\hline                 
Reaction & ID & $\Delta U_{r}$ (kJ mol$^{-1}$) &  $\Delta U_{a}$ (kJ mol$^{-1}$)  \\    
\hline                        
   \ce{C2H2 + H -> C2H3} & AD1.1 & -146.6 [-146.6] & 20.2 [17.8] \\
   \ce{C2H2 + H -> C2H + H2 } & AB1.1 & 122.8 [119.3] & 130.6 [129.8]  \\
   \ce{C2H2 + OH -> C2H2OH} & AD1.2 &-132.4 [-115.7] & 10.1 [15.8]  \\
   \ce{C2H2 + OH -> C2H + H2O} & AB1.2 &78.0 [69.0] & 81.4 [82.1] \\
   \ce{C2H2 + NH2 -> C2H2NH2} & AD1.3 & -107.4 [-89.8] & 29.3 [31.7]  \\
   \ce{C2H2 + NH2 -> C2H + NH3 } & AB1.3 & 114.5 [113.1] & 94.4 [97.0]  \\
   \ce{C2H2 + CH3 -> C2H2CH3} & AD1.4 &-104.3 [-99.6] & 40.7 [42.1]  \\
   \ce{C2H2 + CH3 -> C2H + CH4 } & AB1.4 & 122.1 [122.3]  & 122.9 [122.5]  \\
\hline                                   
\end{tabular}}
\end{table}

Reaction energies and activation barriers for the radical addition and H abstraction are presented in Table \ref{tab:c2h2a}. A visual inspection of the table reveals that DFT values at the MN15-D3BJ/def2-TZVP yield reliable results for the activation energies, with all $\Delta U_{r}$ below or really close to the chemical accuracy of ($\sim$4 kJ mol$^{-1}$).\footnote{1 kJ mol$^{-1}$ = 120.27 K}  All $\Delta U_{r}$ are positive for H abstraction reactions, meaning that all reactions are endothermic, and proceed with very high activation barriers ($\Delta U_{a}$). Therefore, abstraction reactions in ethylene are deemed impossible under astrochemical conditions. In fact, the Kinetic Database for Astrochemistry (KIDA),\footnote{\url{https://kida.astrochem-tools.org}, \citep{Wakelam2012}} reports a rate constant of 0 for AB1.2 in the gas phase in the $\sim$ 10--150 K range. The ethynyl radical (\ce{C2H}) is well known for its reactivity both in the gas phase \citep{Herbst1997, Chastaing1998, Sun2015, Fortenberry2021}, and on ices \citep{Perrero2022}. The transient nature of this radical in comparison with the parent closed shell molecule (\ce{C2H2}) serves as a viable explanation for the elevated endothermicities. In summary, forming the \ce{C2H} is in all cases a non favored process. On the contrary, the addition reactions are in all cases exothermic, with activation barriers that vary between 10.1 kJ mol$^{-1}$ (for OH addition) to 40.7 kJ mol$^{-1}$ (for \ce{CH3} addition). Similarly, the match of our results for AD1.2 is in very good agreement with the literature \citep{Senosiain2005}, with small deviations of 2 kJ mol$^{-1}$ for $\Delta U_{a}$ in our DFT results. However, it is also apparent a difference between our MN15-D3BJ/def2-TZVP and UCCSD(T)-F12/cc-pVTZ-F12//MN15-D3BJ/def2-TZVP values. It is key to make note that \citet{Senosiain2005} reported that coupled cluster was a sub-par method for this particular reaction (and the \ce{OH} addition to \ce{C2H4}, see next section) and that our DFT values may be a better reference in this case. 

\begin{figure}
    \centering
        \includegraphics[width=\linewidth]{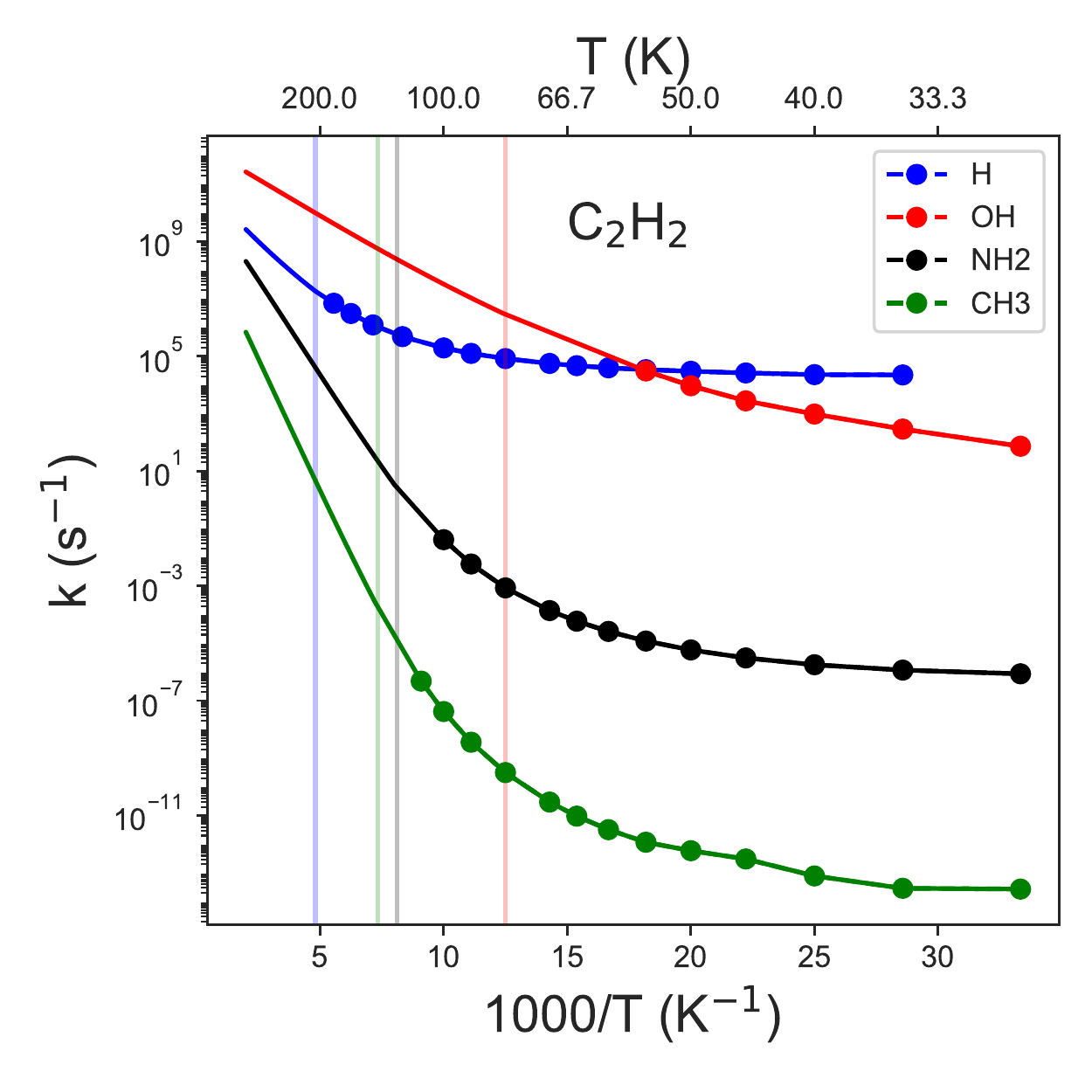} \\
    \caption{Instanton reaction rate constants in the [500,T$_{\text{min}}$] range, with T$_{\text{min}}$ being the lowest temperature achieved in our calculation for each particular reaction starting from \ce{C2H2}. The vertical line represents the crossover temperature, T$_{c}$. }
    \label{fig:ratesc2h2}
\end{figure}

The reaction rate constants for the addition channels are collated in Table \ref{tab:c2h2b}. Instantons were converged to the unimolecular asymptotic limit in most cases, as presented in Figure \ref{fig:ratesc2h2} for the associated rate constants. From the rate constants we observe that reactions for \ce{CH3} and \ce{NH2} are very slow, deeming the reactions non viable under astrophysical conditions. On the contrary, the rate constants for H addition are competitive with H diffusion on ice \citep{Hama2012b, Senevirathne2017, asg17, Nyman2021}. For example, residence times \citep{asg17} span a range that is mostly covered by log$_{10}$k$_{d}$=(-10)--(-3), with our values at $\sim$ -4.50, e.g. in the upper bound of such range. Our results for AD1.2 match very well the previous results presented in \citep{Kobayashi2017}, with rate constants within the order of magnitude. Between both extreme values, we find the reaction with \ce{OH}. Reaction AD1.2 presents a rate constant of k(40 K)=7.3\e{1} s$^{-1}$, which is normally considered a low value when compared with timescales of H diffusion. However, since OH is less mobile, this reaction may compete with OH diffusion. Recent studies are highlighting the importance of the OH radical in surface astrochemistry \citep{Tsuge2021, Ishibashi2021}, and AD1.2 is a good candidate to study these new avenues for reactivity on interstellar ices, due to the small activation barrier. Figure \ref{fig:ratesc2h2} shows the importance of quantum tunneling at low temperatures, rendering the reaction rate constants with H at low temperature orders of magnitude higher than for the heavier radical counterparts in respect to their barriers. This is also nicely reflected by the magnitude of the crossover temperatures presented in Table \ref{fig:ratesc2h2}, because reactions with H present higher transition frequencies, a key magnitude for tunneling.

\begin{table}
\caption{Crossover temperatures (T$_{c}$ in K) and reaction rate-constants k(T$_{\text{min}}$, in s$^{-1}$) the lowest temperature achieved in our calculations (T$_{\text{min}}$ in K) for the exothermic reactions starting from \ce{C2H2}.}             
\label{tab:c2h2b} 
\centering    
\resizebox{\linewidth}{!}{
\begin{tabular}{c c c c  }        
\hline\hline                 
Reaction & $T_{c}$ (K) & T$_{\text{min}}$ (K) & k(T$_{\text{min}}$) (s$^{-1}$)   \\    
\hline                        
\ce{C2H2 + H -> C2H3} & 207 & 35 & 2.2\e{4}  \\
\ce{C2H2 + OH -> C2H2OH} & 80 & 30 & 7.3\e{1} \\
\ce{C2H2 + NH2 -> C2H2NH2} & 123 & 30 & 8.6\e{-7}\\
\ce{C2H2 + CH3 -> C2H2CH3} & 136 & 30 & 2.7\e{-14}  \\
\hline                                   
\end{tabular}}
\end{table}

\subsection{Radicals reacting with ethylene} \label{sec:result_C2H4}

The next molecule for which we investigated abstraction and addition reactions is ethylene (\ce{C2H4}). As in the case of \ce{C2H2}, we studied radical additions and H abstraction reactions, with the main descriptors of the reaction gathered in Table \ref{tab:c2h4a}. A quick visual analysis of the results on the table evinces that once again abstraction reactions are mostly endothermic, with the exception of H abstraction by OH radicals, which is exothermic, and has a moderately low activation barrier. This route can form \ce{C2H3} radicals like reaction AD1.1. All the other abstraction reactions are, as mentioned, endothermic but less than in the previous section. The vinyl radical (\ce{C2H3}) is less reactive than the \ce{C2H} and thus the reactions to form it are less impeded. However, all the endothermic reactions are still non viable in astrophysical environments.

Addition reactions, on the contrary are again always exothermic and exhibit activation barriers in the same order as in the case of the addition reactions presented in the previous section. Once again, for the available comparisons, our values are in good agreement with the available values for H addition (AD2.1), e.g within the order of magnitude for the rate constants \citep{Kobayashi2017} and around $\sim$ 2 kJ mol$^{-1}$ for the value of $\Delta U_{a}$ in AD2.2 \citep{Senosiain2006}. Again, the agreement between coupled-cluster values and DFT ones is slightly worse in the case of reactions involving the OH radical, which is crucial due to the small barriers taking place in reactions with this radical. In \citet{Senosiain2006}, internal tests showed that the coupled cluster method under performed with respect their employed method (UQCISD) that is in much better agreement with our DFT values, increasing the confidence of our subsequent rate constant calculations. For the other heavy radicals \ce{CH3} and \ce{NH2} the barriers are of the same order as for the reactions with \ce{C2H2}, only slightly lower. There is a trend in $\Delta U_{a}$, decreasing with the hybridization of the carbon, check Section \ref{sec:discussion}.

\begin{table}
\caption{Reaction energies ($\Delta U_{r}$ in kJ mol$^{-1}$) and activation energies ($\Delta U_{a}$ in kJ mol$^{-1}$) for the addition and abstraction reactions starting from \ce{C2H4} at the MN15-D3BJ/def2-TZVP level. In square-brackets, values at the UCCSD(T)-F12/cc-pVTZ-F12//MN15-D3BJ/def2-TZVP level of theory.}             
\label{tab:c2h4a} 
\centering                          
\resizebox{\linewidth}{!}{
\begin{tabular}{c c c c }        
\hline\hline                 
Reaction & ID & $\Delta U_{r}$ (kJ mol$^{-1}$) & $\Delta U_{a}$ (kJ mol$^{-1}$)  \\    
\hline                        
   \ce{C2H4 + H -> C2H5} & AD2.1 & -147.2 [-147.2] & 13.5 [11.6]   \\
   \ce{C2H4 + H -> C2H3 + H2} & AB2.1 & 23.8 [23.9]  & 63.0 [62.4]  \\
   \ce{C2H4 + OH -> C2H4OH} & AD2.2 & -115.0 [-104.8]  & 5.9 [11.2] \\ 
   \ce{C2H4 + OH -> C2H3 + H2O} & AB2.2 & -21.6 [-32.3]  & 26.4 [22.5]  \\
   \ce{C2H4 + NH2 -> C2H4NH2} & AD2.3 & -84.4 [-75.8]  & 19.3 [20.8]  \\
   \ce{C2H4 + NH2 -> C2H3 + NH3 } & AB2.3 & 14.1 [16.8] & 59.6 [60.4] \\
   \ce{C2H4 + CH3 -> C2H4CH3 } & AD2.4 & -96.5 [-93.7] & 34.1 [35.1] \\
   \ce{C2H4 + CH3 -> C2H3 + CH4 } & AB2.4 & 23.3 [24.6] & 74.9 [72.4] \\
\hline                                   
\end{tabular}}
\end{table}

\begin{table}
\caption{Crossover temperatures (T$_{c}$ in K) and reaction rate-constants k(T$_{\text{min}}$, in s$^{-1}$) the lowest temperature achieved in our calculations (T$_{\text{min}}$ in K) for the exothermic reactions starting from \ce{C2H4}.}             
\label{tab:c2h4b} 
\centering   
\resizebox{\linewidth}{!}{
\begin{tabular}{c c c c }        
\hline\hline                 
Reaction & $T_{c}$ (K) & T$_{\text{min}}$ (K) & k(T$_{\text{min}}$) (s$^{-1}$)   \\    
\hline                        
\ce{C2H4 + H -> C2H5} & 168 & 40 & 1.2\e{6}  \\
\ce{C2H4 + OH -> C2H4OH} & 49 & 49 & 1.4\e{6},$^{a}$  \\
\ce{C2H4 + OH -> C2H3 + H2O} & 354 & 40 & 5.6\e{1}  \\
\ce{C2H4 + NH2 -> C2H4NH2} & 93 & 30 & 2.6\e{-3} \\
\ce{C2H4 + CH3 -> C2H4CH3} & 121 & 30 & 2.5\e{-12} \\
\hline                                   
\end{tabular}}
\tablefoot{$^{a}$: Instanton rate constants did not converge below $T_{c}$, an Eckart barrier estimate of the rate constant at 30~K is k(30 K)=1.18\e{4} s$^{-1}$.  }
\end{table}

\begin{figure}
    \centering
        \includegraphics[width=\linewidth]{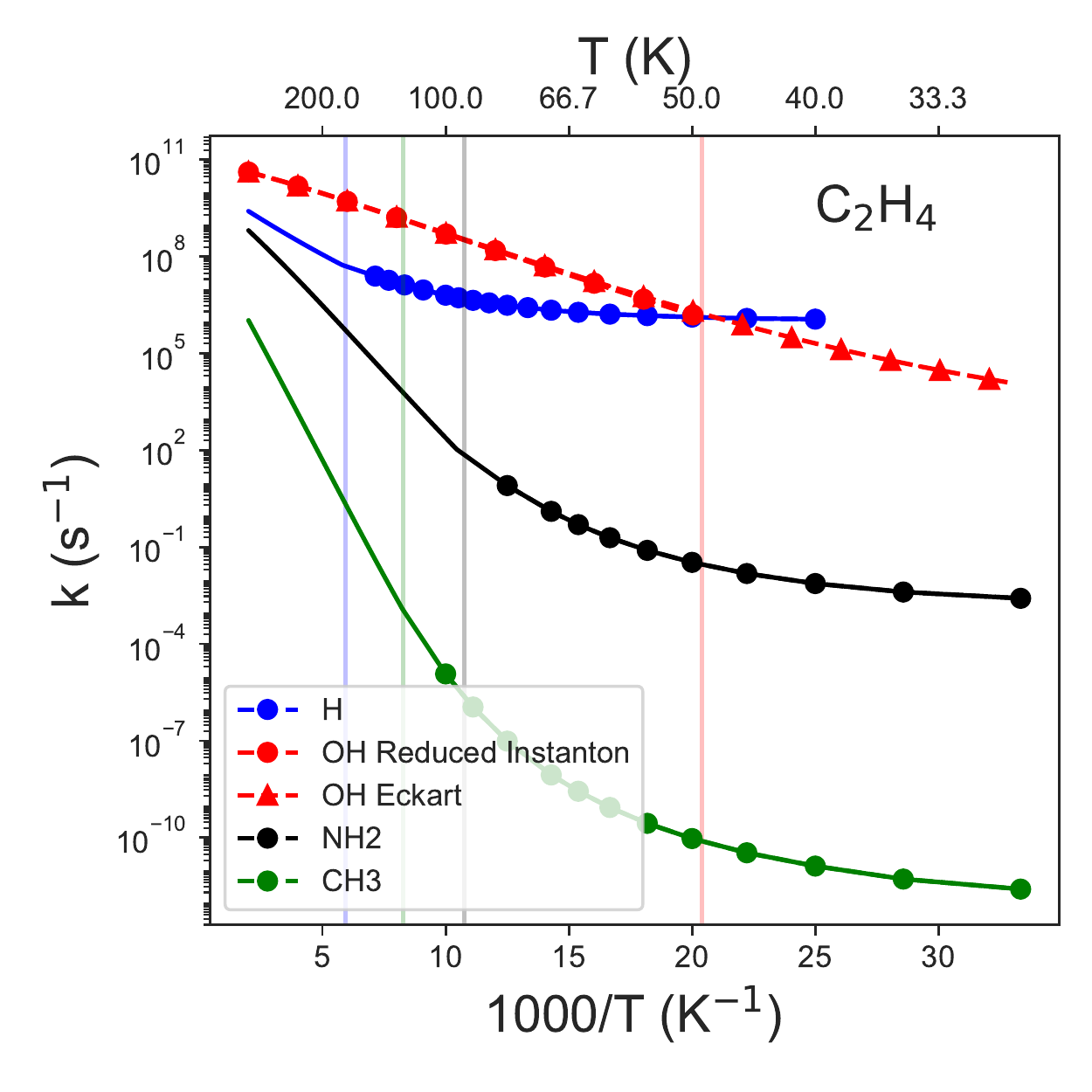} \\
    \caption{Instanton reaction rate constants (and Eckart corrected rate constants in the case of \ce{C2H4 + OH -> C2H4OH}) in the [500,T$_{\text{min}}$] range, with T$_{\text{min}}$ being the lowest temperature achieved in our calculation for each particular reaction for the addition reactions starting from \ce{C2H4}. The vertical line represents the crossover temperature, T$_{c}$. }
    \label{fig:ratesc2h4a}
\end{figure}

\begin{figure}
    \centering
        \includegraphics[width=\linewidth]{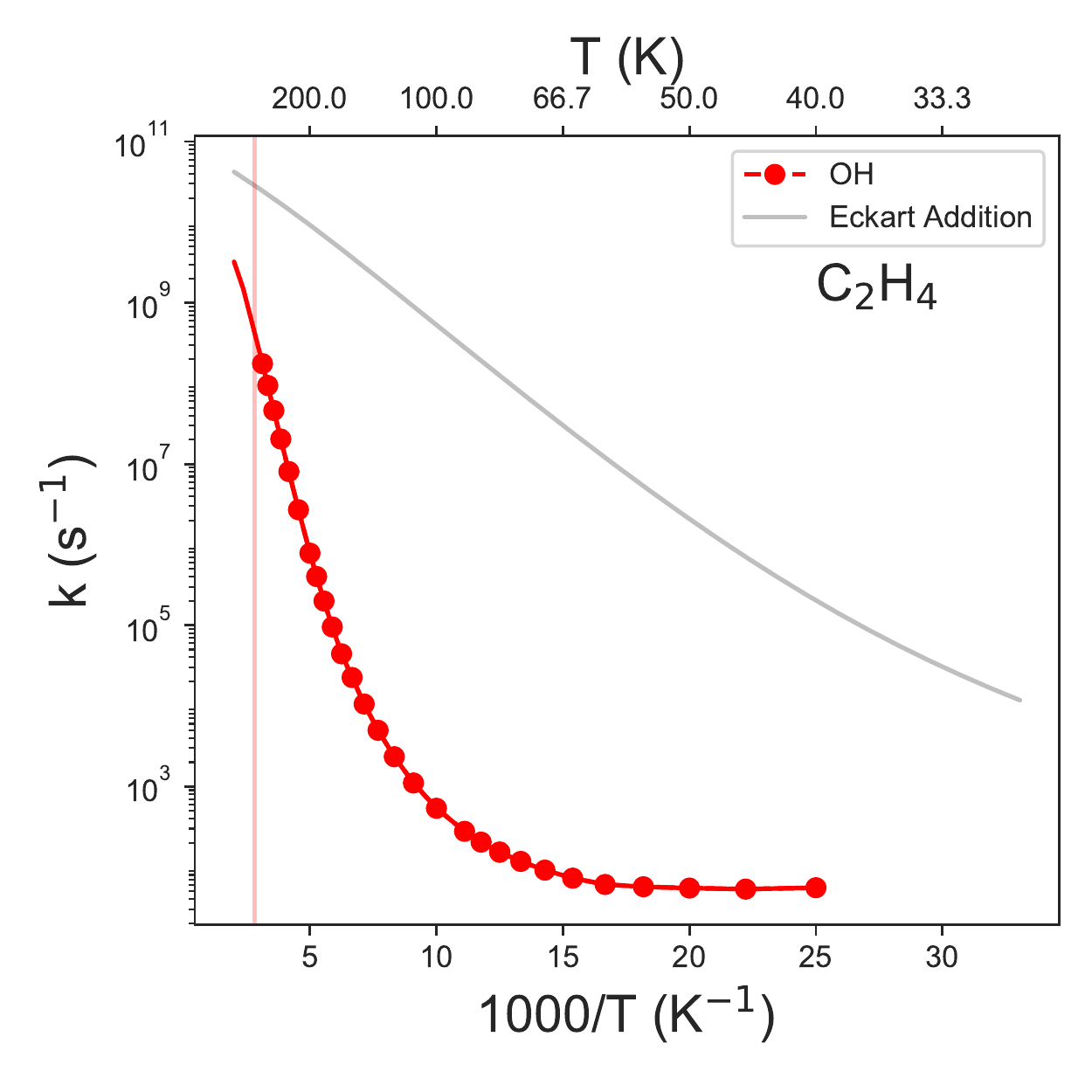} \\
    \caption{Instanton reaction rate constants in the [500,T$_{\text{min}}$] range, with T$_{\text{min}}$ being the lowest temperature achieved in our calculation for the \ce{C2H4 + OH -> C2H3 + H2O}, T$_{c}$. In faded black, reaction rate constants for the \ce{C2H4 + OH -> C2H4OH}. }
    \label{fig:ratesc2h4b}
\end{figure}

All the rate constants for exothermic reactions are gathered in Table \ref{tab:c2h4b} and Figure \ref{fig:ratesc2h4a} for the addition reactions, and Figure \ref{fig:ratesc2h4b} for the abstraction of H by OH. The rate constants confirm the points hinted at by $\Delta U_{a}$. AD2.1 is a fast reaction, also confirmed in \citet{Kobayashi2017}, and is a source of \ce{C2H5} radicals that can react with another radical in a radical-radical recombination. For AD2.2 we were unable to obtain instanton paths below $T_{c}$ and therefore we substituted the instanton rate constants for Eckart corrected rate constants, that we remind, serve as an approximation for the real rate constants. The excellent agreement between reduced instanton rate constants and Eckart in Figure \ref{fig:ratesc2h4a} stems from the value of $T_{c}$, which is small, indicating a lesser influence of quantum effects in this reaction. The Eckart corrected rate constant for AD2.2 is k$^{\text{Eckart}}$(30~K)=1.2\e{4} s$^{-1}$, corresponding to a fast reaction, especially under consideration of the lower mobility of the OH radical on amorphous solid water in comparison with OH. Moreover, and as was introduced before, the OH radical is prone to \emph{non-thermally} diffuse on the surface of the ice, indicating that the addition of \ce{OH} to \ce{C2H4} is a viable reaction under interstellar conditions. 

The abstraction of a H atom from \ce{C2H4} is possible mediated by a OH radical in the AB2.2 reaction. At 40~K the reaction rate constant is 5.6\e{1} s$^{-1}$ and remains constant at lower temperatures, due to the asymptotic behavior, arising from tunneling effects coming purely from the vibrational ground state. On the contrary, the rate constants for the addition reaction do not show an asymptotic behavior in this temperature range, which poses an interesting conundrum. For temperatures close or above 30~K, addition must dominate represented by the two orders of magnitude of difference (see black faded line in \ref{fig:ratesc2h4b}). However, it is unclear, what the behavior at lower temperatures will be, in addition to the error associated with the addition rate constants using an Eckart correction for tunneling. Since both reactions are relatively fast reactions considering the low mobility of OH on water ice we hypothesize that at temperatures T>30~K, which better represent the \emph{non-thermal} regime, addition reactions must dominate. 

\subsection{Radicals reacting with ethane}

\begin{table}
\caption{Reaction energies ($\Delta U_{r}$ in kJ mol$^{-1}$)  and activation energies ($\Delta U_{a}$ in kJ mol$^{-1}$) at the MN15-D3BJ/def2-TZVP level for the abstraction reactions from \ce{C2H6}. In square-brackets, values at the UCCSD(T)-F12/cc-pVTZ-F12//MN15-D3BJ/def2-TZVP level of theory. }             
\label{tab:c2h6a} 
\centering                          
\resizebox{\linewidth}{!}{
\begin{tabular}{c c c c}        
\hline\hline                 
Reaction & ID & $\Delta U_{r}$ & $\Delta U_{a}$   \\    
\hline                        
   \ce{C2H6 + H -> C2H5 + H2 } & AB3.1 & -17.8 [-16.9] & 43.8 [41.3]   \\
   \ce{C2H6 + OH -> C2H5 + H2O } & AB3.2 & -68.1 [-74.0] & 10.4 [9.2]   \\
   \ce{C2H6 + NH2 -> C2H5 + NH3 } & AB3.3 & -28.5 [-26.2] & 45.5 [43.0]   \\
   \ce{C2H6 + CH3 -> C2H5 + CH4 } & AB3.4 & -17.9 [-15.8] &  62.7 [59.1] \\
\hline                                   
\end{tabular}}
\end{table}

To conclude with the ternary of closed shell \ce{C2H_{n}} molecules, we have H abstraction reactions from ethane. These abstractions as opposed to the ones presented in the last sections, are always exothermic and have activation barriers that are lower than for \ce{C2H2} and \ce{C2H4} (See Table \ref{tab:c2h6a}). Logically, addition reactions are precluded in a saturated hydrocarbon such as \ce{C2H6}.

\begin{figure}
    \centering
        \includegraphics[width=\linewidth]{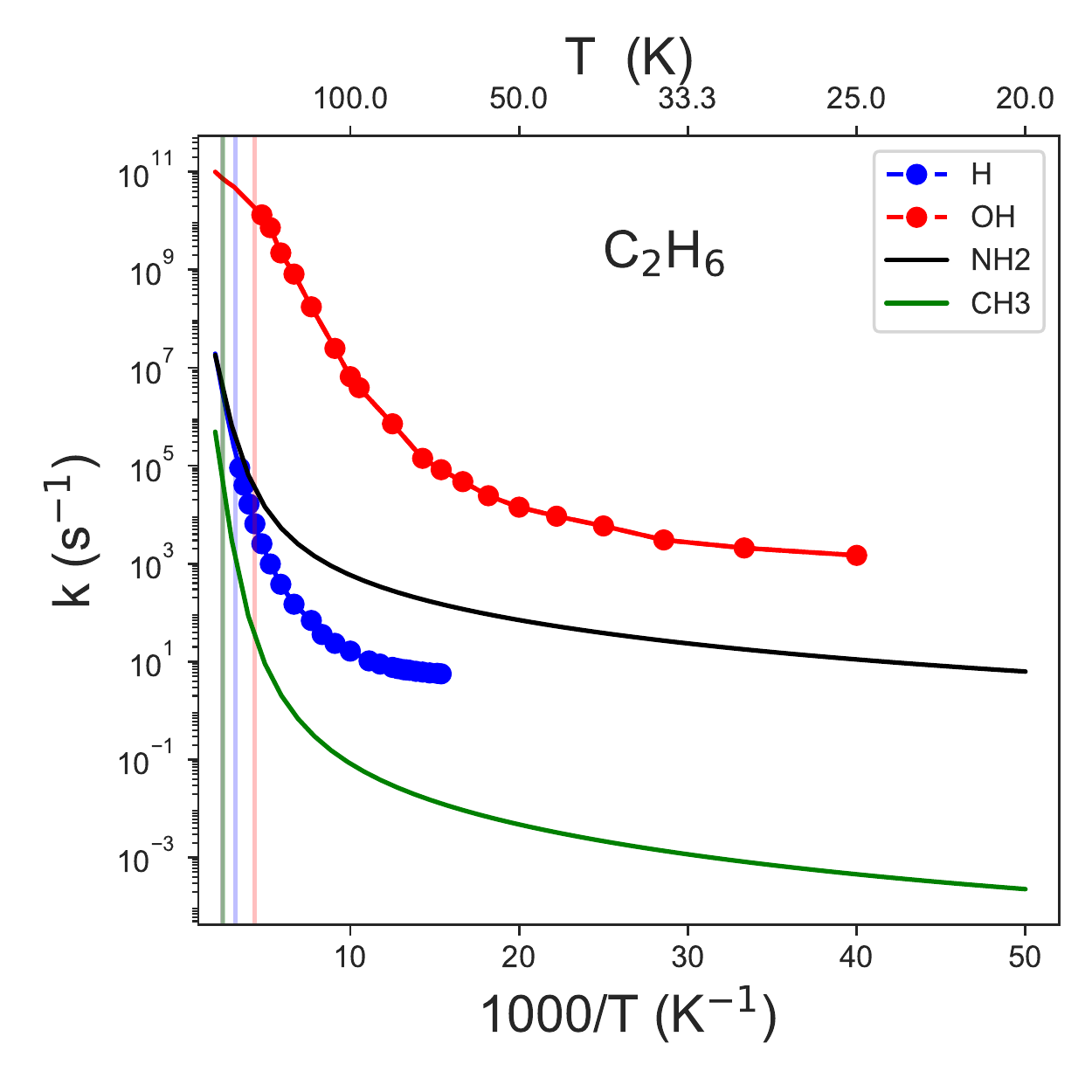} \\
    \caption{Instanton reaction rate constants (and Eckart corrected rate constants in the case of \ce{C2H6 + \ce{NH2}/\ce{CH3} -> \ce{C2H5} + \ce{NH3}/\ce{CH4}}) in the [500,T$_{\text{min}}$] range, with T$_{\text{min}}$ being the lowest temperature achieved in our calculation for each particular reaction for the addition reactions starting from \ce{C2H6}. The vertical line represents the crossover temperature, T$_{c}$. }
    \label{fig:ratesc2h6a}
\end{figure}

The trend in activation energies and reaction energies for this case confirms \ce{OH} as the most reactive agent with $\Delta U_{a}$ tens of kJ mol$^{-1}$ lower than in all the other cases (AB3.2). Similar observations can be done on the basis of $\Delta U_{r}$. For \ce{C2H6}, all reactions present very high $T_{c}$ and therefore, reaction rate constants are significantly affected by quantum tunneling, due to the high transition state frequencies associated with the H-X (X=H, OH, \ce{NH2}, \ce{CH3}) vibrations. The reaction rate constants for this reaction are presented in Table \ref{tab:c2h6b}. Note that, as mentioned during Section \ref{sec:methods}, instanton calculations were particularly difficult for \ce{NH2} and \ce{CH3}, and it was impossible to extract reliable rate constants from them. Hence, we present Eckart corrected rate constants as a compromise.  For abstractions both with \ce{OH} and \ce{NH2}, reaction rate constants are k(25 K)=1.5\e{3} s$^{-1}$ and k(20 K)=6.3\e{0} s$^{-1}$, which are sufficiently high to be viable. However, note that rate constants for AB3.3 are approximated and the here presented results must be adopted with caution. The reliability of the values for \ce{C2H6 + OH -> C2H5 + H2O } is much higher and can be incorporated into astrochemical models. For reactions initiated from \ce{CH3} and \ce{H} the rate constants are either very low (\ce{CH3}) or competitive with diffusion which cast doubts on the viability of these particular routes. A summary of the reaction rate constants is presented in Figure \ref{fig:ratesc2h6a}. All things considered, the formation of the \ce{C2H5} radical is more favorable to proceed from addition reactions in \ce{C2H4} than from the abstraction reactions of this section.

\begin{table}
\caption{Crossover temperatures (T$_{c}$ in K) and reaction rate-constants k(T$_{\text{min}}$, in s$^{-1}$) the lowest temperature achieved in our calculations (T$_{\text{min}}$ in K) for the exothermic reactions starting from \ce{C2H6}.}             
\label{tab:c2h6b} 
\centering                          
\resizebox{\linewidth}{!}{
\begin{tabular}{c c c c}        
\hline\hline                 
Reaction & $T_{c}$ (K) & T$_{\text{min}}$ (K) & k(T$_{\text{min}}$) (s$^{-1}$)    \\    
\hline                        
\ce{C2H6 + H -> C2H5 + H2 } & 312 & 65 & 5.6\e{0}  \\
\ce{C2H6 + OH -> C2H5 + H2O } & 231 & 25 & 1.5\e{3}  \\
\ce{C2H6 + NH2 -> C2H5 + NH3 }$^{a}$ & 412 & 20 &  6.3\e{0}  \\
\ce{C2H6 + CH3 -> C2H5 + CH4 }$^{a}$  & 402 & 20 &   2.3\e{-4}  \\
\hline                                   
\end{tabular}}
\tablefoot{$^{a}$: Rate constants are approximated and obtained from a symmetric Eckart barrier, see text.}
\end{table}

\begin{figure*}
\centering
	\includegraphics[width=8cm]{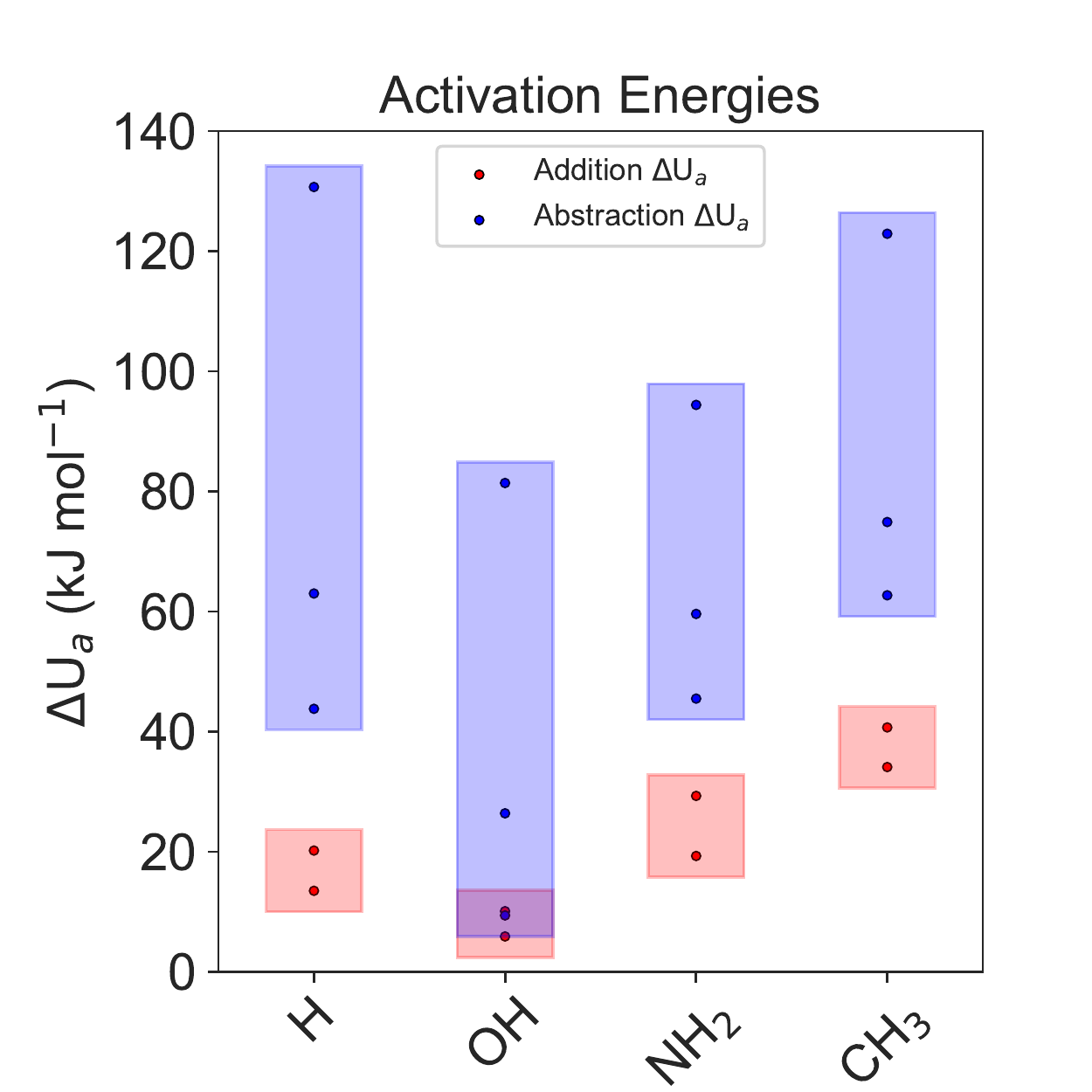} 
	\includegraphics[width=8cm]{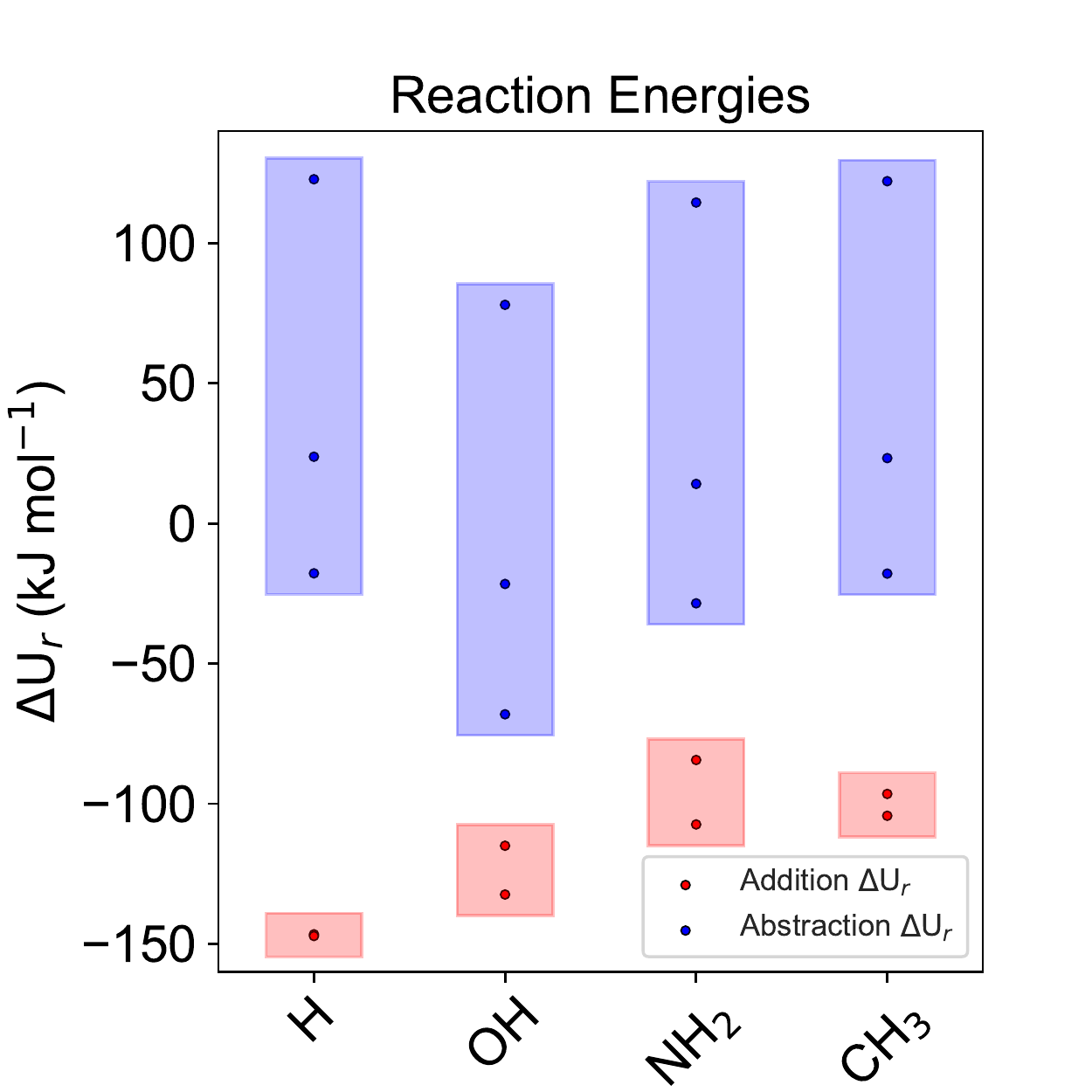}
	\includegraphics[width=8cm]{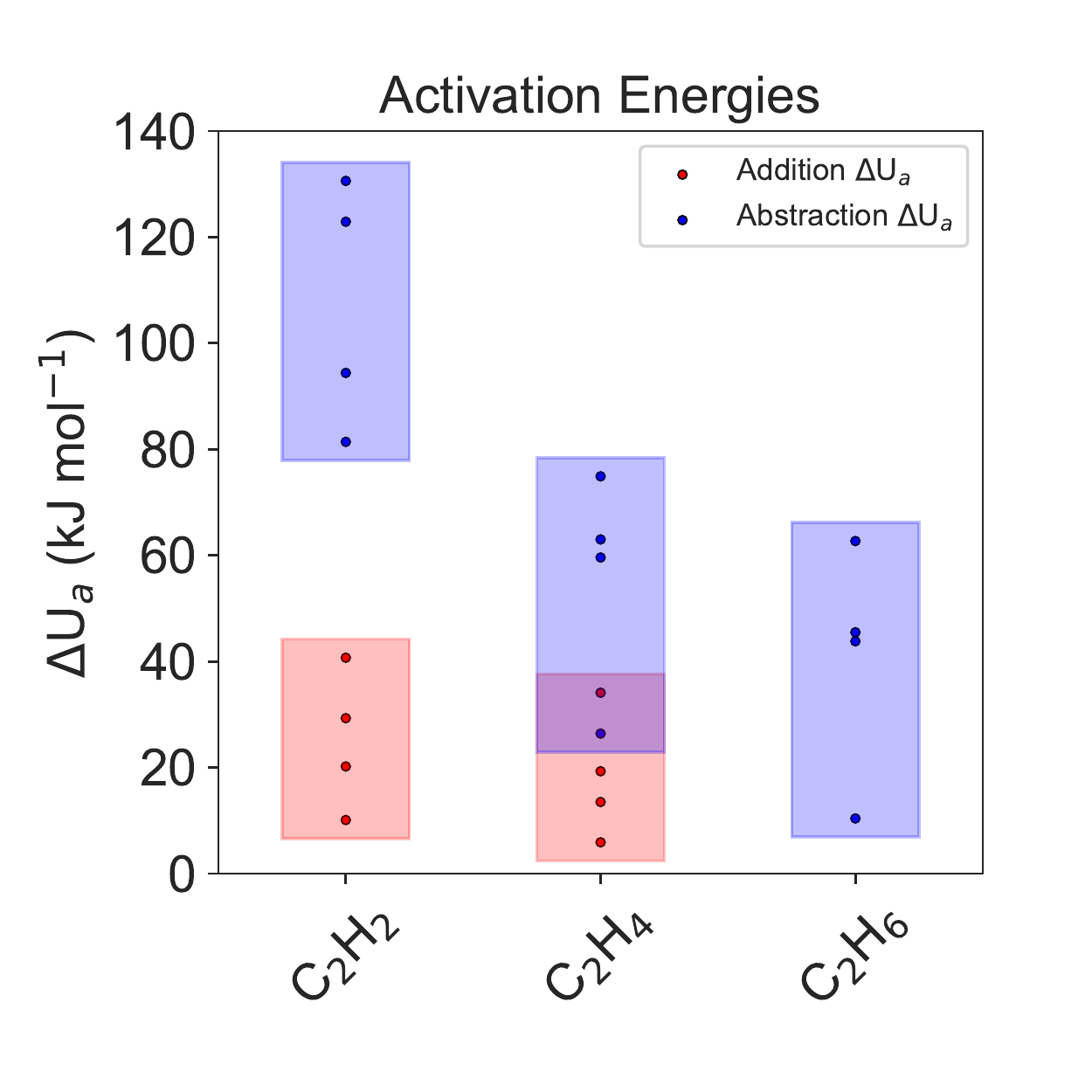} 
	\includegraphics[width=8cm]{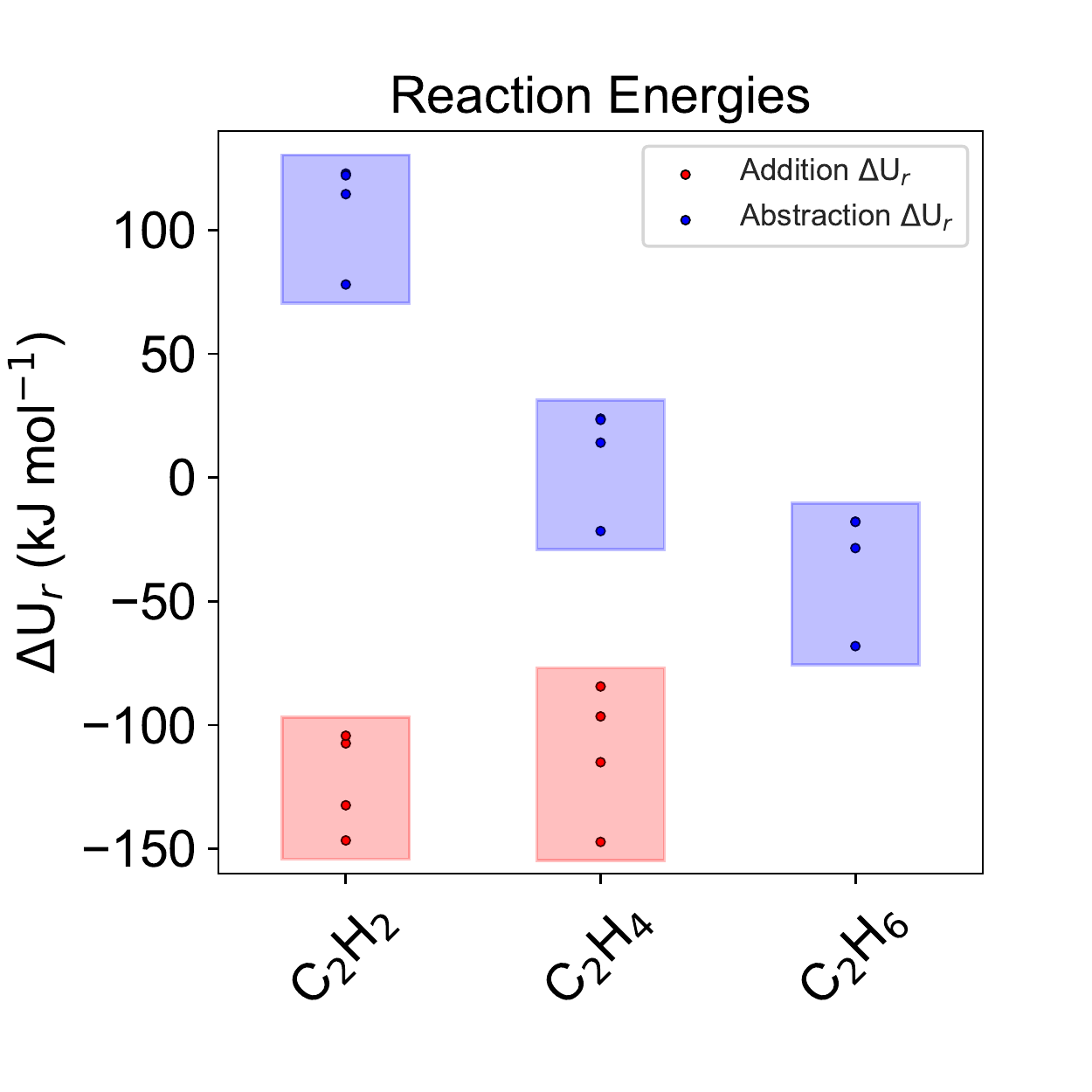}
    \caption{Activation energies ($\Delta U_{a}$ in kJ mol$^{-1}$, left panels) and reaction energies ($\Delta U_{r}$ in kJ mol$^{-1}$, right panels) as a function of the radical under consideration (upper panels) and the reacting hydrocarbon (lower panels).  }
    \label{fig:summary}
\end{figure*}

\section{Discussion} \label{sec:discussion}

We have provided an extensive catalogue of reactions that can help in rationalizing the recent surge of two carbon bearing COMs. Before discussing the astrophysical implications of our work, we would like to bring the attention to the trends that we found, arising from the closed shell molecule under consideration (\ce{C2H2}, \ce{C2H4} and \ce{C2H6}) and the reacting radical (OH, H, \ce{NH2} and \ce{CH3}). The average reaction energies and activation energies found across the sequences H/OH/\ce{NH2}/\ce{CH3} and \ce{C2H2}/\ce{C2H4}/\ce{C2H6} are presented in Figure
\ref{fig:summary}.

The most significant trends that we can observe are summarized as follows. Reaction energies and activation energies mostly decrease following the sequence $\Delta U_{r,a}$(\ce{C2H2}) $\textless$ $\Delta U_{r,a}$(\ce{C2H4}) $\textless$ $\Delta U_{r,a}$(\ce{C2H6}) for abstractions without clear trend for additions. Please note that addition reactions to \ce{C2H6} are forbidden. Following this trend, abstraction reactions start to be exothermic with \ce{C2H4}, with all abstraction reaction in \ce{C2H6} fulfilling this condition. The change in activation energies correlate with the higher exothermicity, with barriers to abstract H from \ce{C2H6} being the lowest of all the abstraction reactions. The reason for the presence of these trends has been indicated before, and it has probably to do with the higher energy of the products. For instance, the ethynil radical (\ce{C2H}) is among the most reactive radicals and the formation of this radical is thermodynamically unfavored from \ce{C2H2}.

We can also extract the trends that appear as a function of the radical partaking in the reaction. By visualizing the top panels of Figure \ref{fig:summary} we observe that the OH radical is the most reactive one by a large margin, presenting the lowest activation barriers of the whole sequence. For these reactions, we also observe that the reaction parameters for \ce{NH2} and \ce{H} are similar, which is also coherent with what was observed in \cite{Markmeyer2019} for H abstraction energies in HC(O)OH (formic acid). Finally, we observe that \ce{CH3} is the most unreactive radical of the ones considered in this work with all reactions yielding in all cases non viable conversions.

From the trends that we observed and the specific values of the rate constants we can infer some important conclusions for the chemistry of COM in cold environments. The key concept arising from this study is that closed shell molecules ``activate" \emph{via} addition and abstraction reactions with (mostly) H and OH. By activating in this context we refer to generate reactive radicals such as \ce{C2H3}, \ce{C2H5} or \ce{C2H4OH} that are prone to further react with other radicals through barrierless radical-radical couplings. It is important to note that radical-radical reactions are not guaranteed to proceed without a barrier (see for example \citet{Enrique-Romero2019}), but are likely to do so in different binding sites. Furthermore, for radical-radical reactions, bimolecular reactions, e.g. proceeding through the Eley-Rideal mechanism are enabled \citep{Ruaud2015}, partially solving the necessary condition of effective surface diffusion imposed by the Langmuir-Hinshelwood one. In this work, we did a systematic search of possible reactions susceptible to occur on top of interstellar dust grains that would produce the previously mentioned radicals. Specifically, the formation of \ce{C2H3} and \ce{C2H5} is favored \emph{via} H addition to \ce{C2H2} and \ce{C2H4} but abstraction reactions from H are not relevant for this mechanism. Heavier radicals, like \ce{CH3} and, to a lesser extent, \ce{NH2} are not significantly reactive neither for addition nor for abstraction. Since tunneling is much less relevant in these cases, moderate barriers (> 15 kJ mol$^{-1}$ or 1800~K) are prohibitive for addition reactions of these radicals. The importance of tunneling is regained again for the reaction AB3.3, and that has a consequential effect in the rate constants for this reaction, probably the only one really viable involving \ce{NH2} from this study. The mentioned reaction is slow nonetheless.

From an astrochemical perspective, the most important finding of the present work is the viability of considering the OH radical an initiator of chemistry on the surface of interstellar dust grains. Reactions on surfaces involving H atoms are usually the most relevant reactions on the surface of dust grains \citep{Hama2013}. However, and very recently, it has been demonstrated that methyl formate can be efficiently formed \emph{via} the photolysis of \ce{H2O}/\ce{CH3OH} ices \citep{Ishibashi2021}. This reaction depends strongly on the photo-dissociation yield of the water ice, indicating that non-thermal \ce{OH} radicals produced in the photolysis are taking part in the reaction. We can establish a parallelism with our results in this study. In water rich environments, where the photolysis of \ce{H2O} proceeds thanks to the secondary UV field \citep{Prasad1983}, OH radicals are formed, with an energy excess that can be used to effectively scan the ice surface \citep{Jin2020}, meeting reaction partners and employing part of this budget to overcome small reaction barriers. With the small barriers found in this work, we suggest that the OH radicals are susceptible of reacting with unsaturated hydrocarbons (\ce{C2H2} and \ce{C2H4}) forming the new species presented above, such as \ce{C2H3}, \ce{C2H5}, \ce{C2H3OH} or \ce{C2H4OH}. Even if the excess energy is only employed in \emph{non-thermal} diffusion, the small barriers for the addition of OH to \ce{C2H2} and \ce{C2H4} provide rate constants that could be overcome.\footnote{Note that in this study we report thermal rate constants and that these just serve as an indication of the timescales for non-thermal events.} It remains to establish the role of the water surface on the whole reaction. A caveat of our simulations is that the effect of a surface is taken into account implicitly. However, calculations performed in similar systems with an explicit account of water molecules (\ce{H2 + OH -> H2O + H}, \citet{Meisner2017}) show that the influence of the water surface in the reaction is small. On a similar note, the OH radical is also effective in generating \ce{C2H5} radicals, \emph{via} reaction AB3.2, affecting also the budget of \ce{C2H5} radicals present on the surface of dust grains available for subsequent reaction to form COMs. For AB2.2, the situation is more complex, owing the competition with the AD2.2 (see Section \ref{sec:result_C2H4}). Experiments are pivotal to understanding what is the leading reaction at 10~K. 

\begin{figure}
\centering
	\includegraphics[width=\linewidth]{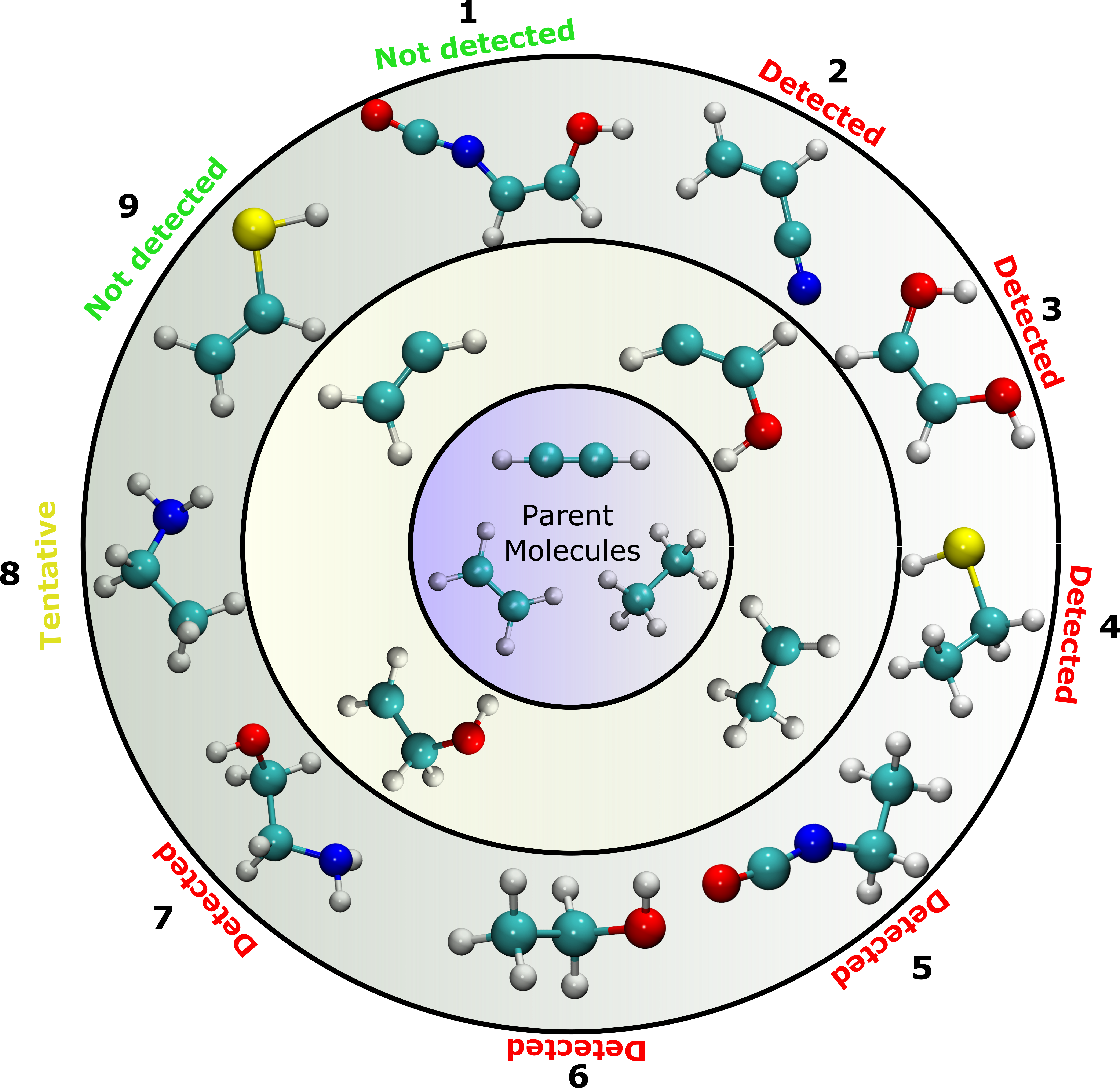} 
    \caption{Schematic diagram showing some of the possible chemical pathways starting from \ce{C2H2}, \ce{C2H4}, and \ce{C2H6} (inner circle) in the formation of COMs \emph{via} activation of the closed shell molecules by OH and H radicals, which forms the four radicals shown in the intermediate circle. Note that not all of the possible reaction paths are indicated in the diagram. The COMs shown in the outer circle are: (1) \ce{CHOHCHNCO}; (2) vinyl cyanide, \ce{C2H3CN}; (3) 1,2-ethenediol, \ce{(CHOH)2}; (4) ethyl mercaptan, \ce{C2H5SH}; (5) ethyl isocyanate \ce{C2H5NCO}; (6) ethanol, \ce{C2H5OH}; (7) ethanolamine, \ce{NH2CH2CH2OH}; (8) ethylamine, \ce{C2H5NH2}; and (9) vinyl mercaptan \ce{C2H3SH}.}
    \label{fig:summary}
\end{figure}

To illustrate the importance of this mechanism in the formation of currently positively, tentatively or not yet detected molecules in the ISM we present a schematic portrait of our suggested mechanism in Figure \ref{fig:summary}. In it, we present several more radicals (i.e. -SH, -NCO, -HCO or -CN)  that the ones explicitly considered in Section \ref{sec:results}, under the consideration that the second radical addition proceeds \emph{via} radical-radical barrierless reactions (see above). From our results, we can confidently say that the radicals \ce{C2H3}, \ce{C2H5}, \ce{C2H2OH}, and \ce{C2H4OH} can be formed \emph{via} reactions AD1.1, AB2.2, AD2.1, AB3.2 AD1.2 and AD2.2. These radicals will stay on the surface waiting for an additional radical to react. In this work we suggest a possible mechanism for the formation of these molecules, in addition to other possibilities such as gas-phase formation, and that can serve to predict new molecules susceptible of detection. Some of these routes were already postulated, as in the case of  \ce{C2H5 + NCO -> C2H5NCO} \citep{Rodriguez-Almeida2021b}. In addition to the extensive list of molecules presented in Section \ref{sec:introduction} there are still non-detections our mechanism may hint at other such as \ce{C2H3NCO} or \ce{C2H3SH}, or tentative detections \ce{C2H5NH2} \citep{Zeng2021}. Similarly, we encourage laboratory studies of molecules falling under this definition, to obtain their rotational spectra, e.g. \ce{HOC2H2CN} and many other arising from the combination of the radicals discussed above.

Another interesting discussion from this work concerns the prevalence of OH over other reactive radicals (in this work represented by \ce{NH2} and \ce{CH3}) may be an explanation for the segregation of O-bearing and N-bearing COMs in cold cores \citep{Jimenez-Serra2021}. Very briefly, and taking as an example the L1498 prestellar core, N- bearing molecules (such as \ce{CH3CN} or \ce{CH2CHCN}) were located in the outer shell of the core, while -O bearing molecules were not and are expected towards the more accreted center of the core.  Nitrogen containing molecules  may be efficiently formed in the gas phase, thanks to the great reactivity of, for example, the CN radical, \citep{Balucani1999, Vazart2015, Sleiman2018, Puzzarini2020} while the formation of oxygen bearing molecules (such as \ce{CH3CHO}, \ce{CH3OCH3}) is favored a later stages, with the depletion of molecules onto interstellar ice surfaces. This hypothesis, in combination with further characterization of hydroxylation reactions atop ices is required to continue deepening into the intricacies of COM formation in interstellar environments.

\section{Conclusion}

In this work, we determined a series of possible routes to form ethyl and vinyl radical and their derivatives, to explain the recent detections of COMs containing at least two carbon atoms. Using quantum chemical calculations we found that the radicals \ce{C2H3}, \ce{C2H5}, \ce{C2H2OH}, and \ce{C2H4OH} are efficiently formed on the surface of grains starting from closed shell molecules and H and OH radicals. The implication of other radicals as \ce{NH2} and \ce{CH3} in the studied reactivity was found to be minor. The here presented reactions act as onsets for additional radical-radical reactions forming COMs. We reiterate the importance of the OH radical in surface processes. The low mobility of the OH radical on interstellar dust grains can be overcome by \emph{non-thermal} effects, as recently studied in the literature \citep{Ishibashi2021}. 

The main caveat of our simulations is that, due to the large amount of reactions under consideration, not all of them are tractable with a homogeneous methodology due to, for example, soft vibrational modes interfering in the convergence of the instantons. Most of the troublesome reactions are not significant for the conclusions our work, but one of them in particular \ce{C2H4 + OH -> C2H4OH} is important, to understand the competition between OH addition and H abstraction in ethylene. Dedicated experiments or calculations (i.e, ring polymer molecular dynamics) are suggested to complete the picture.

We suggest that the mechanism proposed in this work may be behind the recent detection of COMs, and also can be partially responsible of the segregation of N- and O- bearing COMs in prestellar cores. With this work, we would like to encourage further experimental measurements on the viability of further hydroxylation reactions on dust grains, to complement the here presented hypotheses as well as observations of COMs susceptible to be formed by this mechanism in pre-stellar and protostellar cores and in hot cores/corinos.

\begin{acknowledgements}

We thank Prof. Dr. Johannes K\"astner for estimulating discussions. Computer time was granted by the state of Baden-W\"urttemberg through bwHPC and the German Research Foundation (DFG) through grant no. INST 40/467-1FUGG is greatly acknowledged. G.M thanks the Alexander von Humboldt Foundation for a post-doctoral research grant. We thank the Deutsche Forschungsgemeinschaft (DFG, German Research Foundation) for supporting this work by funding EXC 2075 - 390740016 under Germany's Excellence Strategy. We acknowledge the support by the Stuttgart Center for Simulation Science (SimTech).
V.M.R. has received funding from the Comunidad de Madrid through the Atracci\'on de Talento Investigador (Doctores con experiencia) Grant (COOL: Cosmic Origins Of Life; 2019-T1/TIC-15379),  and from the Agencia Estatal de Investigaci\'on (AEI) through the Ram\'on y Cajal programme (grant RYC2020-029387-I).

\end{acknowledgements}

%
%

\bibliographystyle{aa}
\bibliography{sample.bib}

\begin{thebibliography}{72}
\expandafter\ifx\csname natexlab\endcsname\relax\def\natexlab#1{#1}\fi

\bibitem[{{Ag{\'u}ndez} {et~al.}(2021){Ag{\'u}ndez}, {Marcelino}, {Tercero},
  {Cabezas}, {de Vicente}, \& {Cernicharo}}]{Agundez2021}
{Ag{\'u}ndez}, M., {Marcelino}, N., {Tercero}, B., {et~al.} 2021, \aap, 649, L4

\bibitem[{{\'{A}}sgeirsson {et~al.}(2017){\'{A}}sgeirsson, J{\'{o}}nsson, \&
  Wikfeldt}]{asg17}
{\'{A}}sgeirsson, V., J{\'{o}}nsson, H., \& Wikfeldt, K.~T. 2017, Journal of
  Physical Chemistry C, 121, 1648

\bibitem[{Balucani {et~al.}(1999)Balucani, Asvany, Chang, Lin, Lee, Kaiser,
  Bettinger, Schleyer, \& Schaefer}]{Balucani1999}
Balucani, N., Asvany, O., Chang, A.~H., {et~al.} 1999, Journal of Chemical
  Physics, 111, 7457

\bibitem[{{Belloche} {et~al.}(2009){Belloche}, {Garrod}, {M{\"u}ller},
  {Menten}, {Comito}, \& {Schilke}}]{Belloche2009}
{Belloche}, A., {Garrod}, R.~T., {M{\"u}ller}, H.~S.~P., {et~al.} 2009, \aap,
  499, 215

\bibitem[{Cernicharo {et~al.}(2021)Cernicharo, Ag{\'{u}}ndez, Cabezas, \&
  ...}]{Cernicharo2021c}
Cernicharo, J., Ag{\'{u}}ndez, M., Cabezas, C., \& ... 2021, Astron. {\ldots},
  647

\bibitem[{Chastaing {et~al.}(1998)Chastaing, James, Sims, \&
  Smith}]{Chastaing1998}
Chastaing, D., James, P.~L., Sims, I.~R., \& Smith, I.~W. 1998, Faraday
  Discussions, 109, 165

\bibitem[{Chuang {et~al.}(2020)Chuang, Fedoseev, Qasim, Ioppolo, J{\"{a}}ger,
  Henning, Palumbo, {Van Dishoeck}, \& Linnartz}]{Chuang2020}
Chuang, K.~J., Fedoseev, G., Qasim, D., {et~al.} 2020, Astron. Astrophys., 635
  [\eprint[arXiv]{2002.06971}]

\bibitem[{Coleman(1977)}]{col77}
Coleman, S. 1977, Phys. Rev. D, 15, 2929

\bibitem[{Cuppen {et~al.}(2017)Cuppen, Walsh, Lamberts, Semenov, Garrod,
  Penteado, \& Ioppolo}]{Cuppen2017a}
Cuppen, H.~M., Walsh, C., Lamberts, T., {et~al.} 2017, Space Sci. Rev., 212, 1

\bibitem[{Enrique-Romero {et~al.}(2019)Enrique-Romero, Rimola, Ceccarelli,
  Ugliengo, Balucani, \& Skouteris}]{Enrique-Romero2019}
Enrique-Romero, J., Rimola, A., Ceccarelli, C., {et~al.} 2019, ACS Earth Sp.
  Chem., 3, 2158

\bibitem[{Fedoseev {et~al.}(2017)Fedoseev, Chuang, Ioppolo, Qasim, van
  Dishoeck, \& Linnartz}]{Fedoseev2017}
Fedoseev, G., Chuang, K.-J., Ioppolo, S., {et~al.} 2017, Astrophys. J., 842, 52

\bibitem[{Fortenberry(2021)}]{Fortenberry2021}
Fortenberry, R.~C. 2021, The Astrophysical Journal, 921, 132

\bibitem[{Frisch {et~al.}(2016)Frisch, Trucks, Schlegel, Scuseria, Robb,
  Cheeseman, Scalmani, Barone, Petersson, Nakatsuji, Li, Caricato, Marenich,
  Bloino, Janesko, Gomperts, Mennucci, Hratchian, Ortiz, Izmaylov, Sonnenberg,
  Williams-Young, Ding, Lipparini, Egidi, Goings, Peng, Petrone, Henderson,
  Ranasinghe, Zakrzewski, Gao, Rega, Zheng, Liang, Hada, Ehara, Toyota, Fukuda,
  Hasegawa, Ishida, Nakajima, Honda, Kitao, Nakai, Vreven, Throssell,
  Montgomery, Peralta, Ogliaro, Bearpark, Heyd, Brothers, Kudin, Staroverov,
  Keith, Kobayashi, Normand, Raghavachari, Rendell, Burant, Iyengar, Tomasi,
  Cossi, Millam, Klene, Adamo, Cammi, Ochterski, Martin, Morokuma, Farkas,
  Foresman, \& Fox}]{g16}
Frisch, M.~J., Trucks, G.~W., Schlegel, H.~B., {et~al.} 2016, Gaussian 16
  {R}evision {C}.01, gaussian Inc. Wallingford CT

\bibitem[{{Gardner} \& {Winnewisser}(1975)}]{Gardner1975}
{Gardner}, F.~F. \& {Winnewisser}, G. 1975, \apjl, 195, L127

\bibitem[{Garrod {et~al.}(2022)Garrod, Jin, Matis, Jones, Willis, \&
  Herbst}]{Garrod2022}
Garrod, R.~T., Jin, M., Matis, K.~A., {et~al.} 2022, Astrophys. J. Suppl. Ser.,
  259, 1

\bibitem[{Gillan(1987)}]{Gillan1987}
Gillan, M.~J. 1987, Journal of Physics C: Solid State Physics, 20, 3621

\bibitem[{Goerigk {et~al.}(2017)Goerigk, Hansen, Bauer, Ehrlich, Najibi, \&
  Grimme}]{Goerigk2017}
Goerigk, L., Hansen, A., Bauer, C., {et~al.} 2017, Phys. Chem. Chem. Phys., 19,
  32184

\bibitem[{Grimme {et~al.}(2011)Grimme, Ehrlich, \& Goerigk}]{Grimme2011}
Grimme, S., Ehrlich, S., \& Goerigk, L. 2011, J. Comp. Chem., 32, 1456,
  https://doi.org/10.1002/jcc.21759

\bibitem[{Hama {et~al.}(2012)Hama, Kuwahata, Watanabe, Kouchi, Kimura, Chigai,
  \& Pirronello}]{Hama2012b}
Hama, T., Kuwahata, K., Watanabe, N., {et~al.} 2012, Astrophys. J., 757, 185

\bibitem[{Hama \& Watanabe(2013)}]{Hama2013}
Hama, T. \& Watanabe, N. 2013, Chem. Rev., 113, 8783

\bibitem[{Herbst \& {Van Dishoeck}(2009)}]{Herbst2009}
Herbst, E. \& {Van Dishoeck}, E.~F. 2009, Annual Review of Astronomy and
  Astrophysics, 47, 427

\bibitem[{Herbst \& Woon(1997)}]{Herbst1997}
Herbst, E. \& Woon, D.~E. 1997, The Astrophysical Journal, 489, 109

\bibitem[{Hollis {et~al.}(2004)Hollis, Jewell, Lovas, Remijan, \&
  M{\o}llendal}]{Hollis2004}
Hollis, J.~M., Jewell, P.~R., Lovas, F.~J., Remijan, A., \& M{\o}llendal, H.
  2004, The Astrophysical Journal, 610, L21

\bibitem[{Ioppolo {et~al.}(2021)Ioppolo, Fedoseev, Chuang, Cuppen, Clements,
  Jin, Garrod, Qasim, Kofman, van Dishoeck, \& Linnartz}]{Ioppolo2020}
Ioppolo, S., Fedoseev, G., Chuang, K.~J., {et~al.} 2021, Nat. Astron., 5, 197

\bibitem[{Ishibashi {et~al.}(2021)Ishibashi, Hidaka, Oba, Kouchi, \&
  Watanabe}]{Ishibashi2021}
Ishibashi, A., Hidaka, H., Oba, Y., Kouchi, A., \& Watanabe, N. 2021, The
  Astrophysical Journal Letters, 921, L13

\bibitem[{Jim{\'{e}}nez-Serra {et~al.}(2021)Jim{\'{e}}nez-Serra, Vasyunin,
  Spezzano, Caselli, Cosentino, \& Viti}]{Jimenez-Serra2021}
Jim{\'{e}}nez-Serra, I., Vasyunin, A.~I., Spezzano, S., {et~al.} 2021, The
  Astrophysical Journal, 917, 44

\bibitem[{Jin \& Garrod(2020)}]{Jin2020}
Jin, M. \& Garrod, R.~T. 2020, Astrophys. J. Suppl. Ser., 249, 26

\bibitem[{{Johnson} {et~al.}(1977){Johnson}, {Lovas}, {Gottlieb}, {Gottlieb},
  {Litvak}, {Guelin}, \& {Thaddeus}}]{Johnson1977}
{Johnson}, D.~R., {Lovas}, F.~J., {Gottlieb}, C.~A., {et~al.} 1977, \apj, 218,
  370

\bibitem[{K\"astner {et~al.}(2009)K\"astner, Carr, Keal, Thiel, Wander, \&
  Sherwood}]{kae09a}
K\"astner, J., Carr, J.~M., Keal, T.~W., {et~al.} 2009, J. Phys. Chem. A, 113,
  11856

\bibitem[{{Kelvin Lee} {et~al.}(2021){Kelvin Lee}, Loomis, Burkhardt, Cooke,
  Xue, Siebert, Shingledecker, Remijan, Charnley, McCarthy, \&
  McGuire}]{KelvinLee2021}
{Kelvin Lee}, K.~L., Loomis, R.~A., Burkhardt, A.~M., {et~al.} 2021, Astrophys.
  J. Lett., 908, L11

\bibitem[{Knizia {et~al.}(2009)Knizia, Adler, \& Werner}]{Knizia2009}
Knizia, G., Adler, T.~B., \& Werner, H.-J. 2009, The Journal of Chemical
  Physics, 130, 054104

\bibitem[{Kobayashi {et~al.}(2017)Kobayashi, Hidaka, Lamberts, Hama, Kawakita,
  K{\"{a}}stner, \& Watanabe}]{Kobayashi2017}
Kobayashi, H., Hidaka, H., Lamberts, T., {et~al.} 2017, The Astrophysical
  Journal, 837, 155

\bibitem[{Lamberts {et~al.}(2016)Lamberts, Samanta, Köhn, \&
  Kästner}]{Lamberts2016}
Lamberts, T., Samanta, P.~K., Köhn, A., \& Kästner, J. 2016, Phys. Chem.
  Chem. Phys., 18, 33021

\bibitem[{Langer(1967)}]{lan67}
Langer, J.~S. 1967, Ann. Phys. (N.Y.), 41, 108

\bibitem[{Markmeyer {et~al.}(2019)Markmeyer, Lamberts, Meisner, \&
  K{\"{a}}stner}]{Markmeyer2019}
Markmeyer, M.~N., Lamberts, T., Meisner, J., \& K{\"{a}}stner, J. 2019, Mon.
  Not. Roy. Astron. Soc., 482, 293

\bibitem[{McConnell \& K{\"{a}}stner(2017)}]{McConnell2017}
McConnell, S. \& K{\"{a}}stner, J. 2017, J. Comp. Chem., 38, 2570

\bibitem[{McGuire(2022)}]{Mcguire2022}
McGuire, B.~A. 2022, Astrophys. J. Suppl. Ser., 259, 30

\bibitem[{Meisner {et~al.}(2017)Meisner, Lamberts, \& K\"astner}]{Meisner2017}
Meisner, J., Lamberts, T., \& K\"astner, J. 2017, ACS Earth Space Chem., 1, 399

\bibitem[{Metz {et~al.}(2014)Metz, K{\"{a}}stner, Sokol, Keal, \&
  Sherwood}]{Chemshell}
Metz, S., K{\"{a}}stner, J., Sokol, A.~A., Keal, T.~W., \& Sherwood, P. 2014,
  Wiley Interdiscip. Rev. Comput. Mol. Sci., 4, 101

\bibitem[{Miksch {et~al.}(2021)Miksch, Riffelt, Oliveira, Kästner, \&
  Molpeceres}]{Miksch2021}
Miksch, A.~M., Riffelt, A., Oliveira, R., Kästner, J., \& Molpeceres, G. 2021,
  Monthly Notices of the Royal Astronomical Society, 505, 3157

\bibitem[{Miller(1975)}]{mil75}
Miller, W.~H. 1975, J. Chem. Phys., 62, 1899

\bibitem[{Molpeceres {et~al.}(2021)Molpeceres, de~la Concepción, \&
  Serra}]{molpeceres2021diastereoselective}
Molpeceres, G., de~la Concepción, J.~G., \& Serra, I.~J. 2021,
  Diastereoselective Formation of trans-HC(O)SH Through Hydrogenation of OCS on
  Interstellar Dust Grains

\bibitem[{Molpeceres \& K{\"{a}}stner(2021)}]{Molpeceres2021}
Molpeceres, G. \& K{\"{a}}stner, J. 2021, Astrophys. J., 910, 55

\bibitem[{Molpeceres \& Rivilla(2022)}]{molpeceres_2022_6581077}
Molpeceres, G. \& Rivilla, V. 2022, {Cartesian coordinates and energetic
  parameters described in the present manuscript,
  \url{https://doi.org/10.5281/zenodo.6581077}}

\bibitem[{Nyman(2021)}]{Nyman2021}
Nyman, G. 2021, Frontiers in Astronomy and Space Sciences, 8

\bibitem[{Perrero {et~al.}(2022)Perrero, Enrique-Romero, Mart{\'{i}}nez-Bachs,
  Ceccarelli, Balucani, Ugliengo, \& Rimola}]{Perrero2022}
Perrero, J., Enrique-Romero, J., Mart{\'{i}}nez-Bachs, B., {et~al.} 2022, ACS
  Earth and Space Chemistry, 6, 496

\bibitem[{Prasad \& Tarafdar(1983)}]{Prasad1983}
Prasad, S.~S. \& Tarafdar, S.~P. 1983, ApJ, 267, 603

\bibitem[{Puzzarini(2022)}]{Puzzarini2022}
Puzzarini, C. 2022, Front. Astron. Sp. Sci., 8

\bibitem[{Puzzarini {et~al.}(2020)Puzzarini, Salta, Tasinato, Lupi, Cavallotti,
  \& Barone}]{Puzzarini2020}
Puzzarini, C., Salta, Z., Tasinato, N., {et~al.} 2020, Monthly Notices of the
  Royal Astronomical Society, 496, 4298

\bibitem[{Quan {et~al.}(2016)Quan, Herbst, Corby, Durr, \& Hassel}]{Quan2016}
Quan, D., Herbst, E., Corby, J.~F., Durr, A., \& Hassel, G. 2016, Astrophys.
  J., 824, 129

\bibitem[{{Rivilla} {et~al.}(2022){Rivilla}, {Colzi}, {Jim{\'e}nez-Serra},
  {Mart{\'\i}n-Pintado}, {Meg{\'\i}as}, {Melosso}, {Bizzocchi},
  {L{\'o}pez-Gallifa}, {Mart{\'\i}nez-Henares}, {Massalkhi}, {Tercero}, {de
  Vicente}, {Guillemin}, {Garc{\'\i}a de la Concepci{\'o}n}, {Rico-Villas},
  {Zeng}, {Mart{\'\i}n}, {Requena-Torres}, {Tonolo}, {Alessandrini}, {Dore},
  {Barone}, \& {Puzzarini}}]{Rivilla2022}
{Rivilla}, V.~M., {Colzi}, L., {Jim{\'e}nez-Serra}, I., {et~al.} 2022, \apjl,
  929, L11

\bibitem[{Rivilla {et~al.}(2021)Rivilla, Jiménez-Serra, Martín-Pintado,
  Briones, Rodríguez-Almeida, Rico-Villas, Tercero, Zeng, Colzi, de~Vicente,
  Martín, \& Requena-Torres}]{Rivilla2021}
Rivilla, V.~M., Jiménez-Serra, I., Martín-Pintado, J., {et~al.} 2021,
  Proceedings of the National Academy of Sciences, 118, e2101314118

\bibitem[{Rodr{\'{\i}}guez-Almeida {et~al.}(2021)Rodr{\'{\i}}guez-Almeida,
  Jim{\'{e}}nez-Serra, Rivilla, Mart{\'{\i}}n-Pintado, Zeng, Tercero,
  de~Vicente, Colzi, Rico-Villas, Mart{\'{\i}}n, \&
  Requena-Torres}]{RodriguezAlmeida2021}
Rodr{\'{\i}}guez-Almeida, L.~F., Jim{\'{e}}nez-Serra, I., Rivilla, V.~M.,
  {et~al.} 2021, The Astrophysical Journal Letters, 912, L11

\bibitem[{{Rodr{\'\i}guez-Almeida} {et~al.}(2021){Rodr{\'\i}guez-Almeida},
  {Rivilla}, {Jim{\'e}nez-Serra}, {Melosso}, {Colzi}, {Zeng}, {Tercero}, {de
  Vicente}, {Mart{\'\i}n}, {Requena-Torres}, {Rico-Villas}, \&
  {Mart{\'\i}n-Pintado}}]{Rodriguez-Almeida2021b}
{Rodr{\'\i}guez-Almeida}, L.~F., {Rivilla}, V.~M., {Jim{\'e}nez-Serra}, I.,
  {et~al.} 2021, \aap, 654, L1

\bibitem[{Rommel {et~al.}(2011)Rommel, Goumans, \& K\"astner}]{Rommel2011-2}
Rommel, J.~B., Goumans, T.~P., \& K\"astner, J. 2011, J. Chem. Theory Comp., 7,
  690

\bibitem[{Rommel \& K{\"{a}}stner(2011)}]{Rommel2011}
Rommel, J.~B. \& K{\"{a}}stner, J. 2011, J. Chem. Phys., 134, 184107

\bibitem[{Ruaud {et~al.}(2015)Ruaud, Loison, Hickson, Gratier, Hersant, \&
  Wakelam}]{Ruaud2015}
Ruaud, M., Loison, J.~C., Hickson, K.~M., {et~al.} 2015, Monthly Notices of the
  Royal Astronomical Society, 447, 4004

\bibitem[{Senevirathne {et~al.}(2017)Senevirathne, Andersson, Dulieu, \&
  Nyman}]{Senevirathne2017}
Senevirathne, B., Andersson, S., Dulieu, F., \& Nyman, G. 2017, Mol.
  Astrophys., 6, 59

\bibitem[{Senosiain {et~al.}(2005)Senosiain, Klippenstein, \&
  Miller}]{Senosiain2005}
Senosiain, J.~P., Klippenstein, S.~J., \& Miller, J.~A. 2005, The Journal of
  Physical Chemistry A, 109, 6045

\bibitem[{Senosiain {et~al.}(2006)Senosiain, Klippenstein, \&
  Miller}]{Senosiain2006}
Senosiain, J.~P., Klippenstein, S.~J., \& Miller, J.~A. 2006, Journal of
  Physical Chemistry A, 110, 6960

\bibitem[{Sleiman {et~al.}(2018)Sleiman, {El Dib}, Rosi, Skouteris, Balucani,
  \& Canosa}]{Sleiman2018}
Sleiman, C., {El Dib}, G., Rosi, M., {et~al.} 2018, Physical Chemistry Chemical
  Physics, 20, 5478

\bibitem[{Sun {et~al.}(2015)Sun, Huang, \& Lee}]{Sun2015}
Sun, Y.~L., Huang, W.~J., \& Lee, S.~H. 2015, Journal of Physical Chemistry
  Letters, 6, 4117

\bibitem[{{Tercero} {et~al.}(2018){Tercero}, {Cuadrado}, {L{\'o}pez},
  {Brouillet}, {Despois}, \& {Cernicharo}}]{Tercero2018}
{Tercero}, B., {Cuadrado}, S., {L{\'o}pez}, A., {et~al.} 2018, \aap, 620, L6

\bibitem[{Tsuge \& Watanabe(2021)}]{Tsuge2021}
Tsuge, M. \& Watanabe, N. 2021, Accounts of Chemical Research, 54, 471

\bibitem[{Vazart {et~al.}(2015)Vazart, Latouche, Skouteris, Balucani, \&
  Barone}]{Vazart2015}
Vazart, F., Latouche, C., Skouteris, D., Balucani, N., \& Barone, V. 2015,
  Astrophy. J., 810, 111

\bibitem[{Wakelam {et~al.}(2017{\natexlab{a}})Wakelam, Bron, Cazaux, Dulieu,
  Gry, Guillard, Habart, Hornek{\ae}r, Morisset, Nyman, Pirronello, Price,
  Valdivia, Vidali, \& Watanabe}]{Wakelam2017-b}
Wakelam, V., Bron, E., Cazaux, S., {et~al.} 2017{\natexlab{a}}, Molecular
  Astrophysics, 9, 1

\bibitem[{Wakelam {et~al.}(2012)Wakelam, Herbst, Loison, Smith, Chandrasekaran,
  Pavone, Adams, Bacchus-Montabonel, Bergeat, B{\'{e}}roff, Bierbaum, Chabot,
  Dalgarno, van Dishoeck, Faure, Geppert, Gerlich, Galli, H{\'{e}}brard,
  Hersant, Hickson, Honvault, Klippenstein, Picard, Nyman, Pernot, Schlemmer,
  Selsis, Sims, Talbi, Tennyson, Troe, Wester, \& Wiesenfeld}]{Wakelam2012}
Wakelam, V., Herbst, E., Loison, J.-C., {et~al.} 2012, The Astrophysical
  Journal Supplement Series, 199, 21

\bibitem[{Wakelam {et~al.}(2017{\natexlab{b}})Wakelam, Loison, Mereau, \&
  Ruaud}]{Wakelam2017b}
Wakelam, V., Loison, J.~C., Mereau, R., \& Ruaud, M. 2017{\natexlab{b}}, Mol.
  Astrophy., 6, 22

\bibitem[{Weigend \& Ahlrichs(2005)}]{Weigend2005}
Weigend, F. \& Ahlrichs, R. 2005, Phys. Chem. Chem. Phys., 7, 3297

\bibitem[{Yu {et~al.}(2016)Yu, He, Li, \& Truhlar}]{Yu2016}
Yu, H.~S., He, X., Li, S.~L., \& Truhlar, D.~G. 2016, Chem. Sci., 7, 5032

\bibitem[{{Zeng} {et~al.}(2021){Zeng}, {Jim{\'e}nez-Serra}, {Rivilla},
  {Mart{\'\i}n-Pintado}, {Rodr{\'\i}guez-Almeida}, {Tercero}, {de Vicente},
  {Rico-Villas}, {Colzi}, {Mart{\'\i}n}, \& {Requena-Torres}}]{Zeng2021}
{Zeng}, S., {Jim{\'e}nez-Serra}, I., {Rivilla}, V.~M., {et~al.} 2021, \apjl,
  920, L27

\bibitem[{{Zuckerman} {et~al.}(1975){Zuckerman}, {Turner}, {Johnson}, {Clark},
  {Lovas}, {Fourikis}, {Palmer}, {Morris}, {Lilley}, {Ball}, {Gottlieb},
  {Litvak}, \& {Penfield}}]{Zuckerman1975}
{Zuckerman}, B., {Turner}, B.~E., {Johnson}, D.~R., {et~al.} 1975, \apjl, 196,
  L99

\end{thebibliography}

\begin{appendix}
\section{Test of implicit surface approach} \label{sec:appendix}

\begin{figure}
\centering
	\includegraphics[width=8cm]{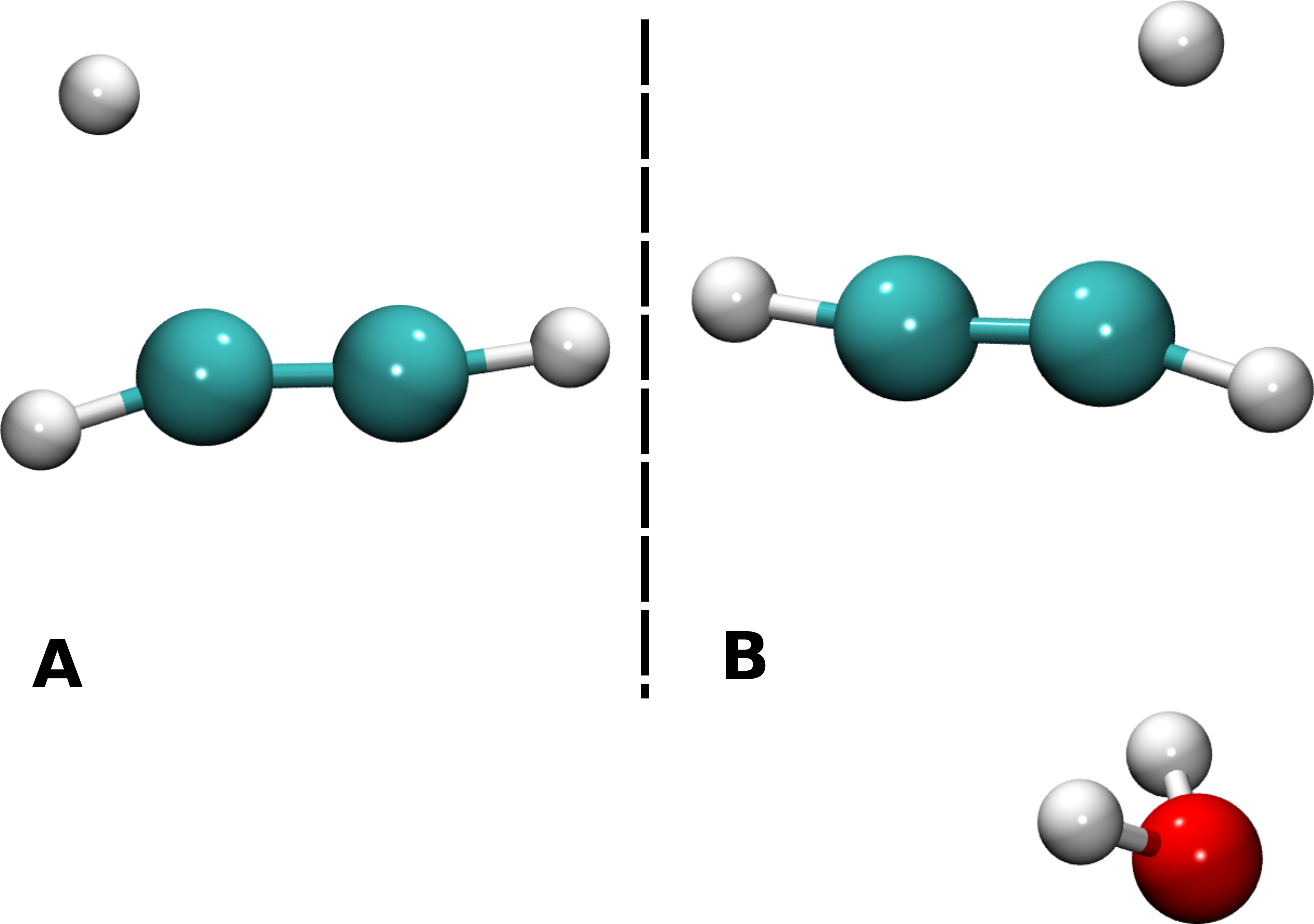}
    \caption{Models for the test of the implicit (A) \emph{vs} explicit surface approach (B). The depicted figures correspond to reaction AD1.1}
    \label{fig:example_appendix}
\end{figure}

Our calculations use an implicit surface approach to account for surface effects in the simulations. In such approach, which was validated for weak and intermediate binding situations \citep{Meisner2017, Lamberts2016, molpeceres2021diastereoselective}, the rotational contributions to the partition function are constant and consequentially their effects in the rate constants are cancelled out. This vanishes the effect of the rotational thermal energy and entropy in the rate constant calculation. We have updated the consistency tests for the application of this approximation in this article. In particular, we have compared the bimolecular activation energies including ZPE ($\Delta U_{a}$), e.g. with the energy of the reactants calculated separately (note that in the main text we describe unimolecular activation energies) for a gas-phase model and the single water molecule depicted in \cite{Wakelam2017-b} for \ce{C2H2}, \ce{C2H4} and \ce{C2H6} (re-optimized at the MN15-D3BJ/def2-TZVP level of theory). We tested the validity of the approach for reactions AD1.1, AD2.1 and AB3.1 of the main text. In Figure \ref{fig:example_appendix} we show a graphical example of how both models look. The results for the validation calculations are gathered in Table \ref{tab:benchmark}. All the differences are below 1 kJ mol$^{-1}$ , indicating a tiny influence of the water molecule in the reaction activation energy. 

\begin{table}
\caption{Comparison between the activation energies including ZPE for the implicit surface approach ($\Delta U_{a, imp}$) \emph{vs} a model explicitly including one water molecule ($\Delta U_{a, exp}$).}             
\label{tab:benchmark} 
\centering                          
\resizebox{\linewidth}{!}{
\begin{tabular}{c c c }        
\hline\hline                 
Reaction &  $\Delta U_{a, imp}$ (kJ mol$^{-1}$) & $\Delta U_{a, exp}$ (kJ mol$^{-1}$)  \\    
\hline                        
AD1.1 & 20.8 & 21.2           \\
AD2.1 & 14.1 & 15.0         \\
AB3.1 & 44.9 & 44.5   \\
\hline                                   
\end{tabular}}
\end{table}

\end{appendix}

\end{document}